\documentclass[sigconf]{acmart}

\AtBeginDocument{%
  \providecommand\BibTeX{{%
    \normalfont B\kern-0.5em{\scshape i\kern-0.25em b}\kern-0.8em\TeX}}}

\setcopyright{rightsretained}

\copyrightyear{2026}
\acmYear{2026}
\setcopyright{cc}
\setcctype{by}
\acmConference[CHI '26]{Proceedings of the 2026 CHI Conference on Human Factors in Computing Systems}{April 13--17, 2026}{Barcelona, Spain}
\acmBooktitle{Proceedings of the 2026 CHI Conference on Human Factors in Computing Systems (CHI '26), April 13--17, 2026, Barcelona, Spain}
\acmPrice{}
\acmDOI{10.1145/3772318.3790375}
\acmISBN{979-8-4007-2278-3/2026/04}

\usepackage{multirow}
\usepackage{subcaption}
\usepackage{caption}
\usepackage[normalem]{ulem}
\usepackage{siunitx}
\usepackage{booktabs}
\usepackage{array}
\usepackage{tabularx}
\usepackage{xcolor}
\usepackage{multicol}
\usepackage{siunitx}
\usepackage{pifont}
\usepackage[utf8]{inputenc}

\newcommand{\SharedGrayBox}[2]{%
  \par\vspace{4pt}
  \noindent
  \hspace*{-6pt}
  \begingroup
  \setlength{\fboxsep}{5pt}      
  \setlength{\fboxrule}{1pt}      
  \fcolorbox{gray!75!black}{gray!5}{%
    \begin{minipage}{0.98\linewidth}
      \textbf{#1}\par
      \vspace{6pt}
      {\ttfamily\footnotesize
      #2\par
      }
    \end{minipage}
  }%
  \endgroup
  \par\vspace{4pt}
}

\begin{document}

\title[Towards AI as Colleagues: Multi-Agent System Improves Structured Ideation Processes]{Towards AI as Colleagues:  \\ Multi-Agent System Improves Structured Ideation Processes}

\author{Kexin Quan}
\affiliation{%
  \institution{Information Sciences, \\University of Illinois Urbana-Champaign}
  \city{Champaign}
  \state{Illinois}
  \country{USA}}
\email{kq4@illinois.edu}

\author{Dina Albassam}
\authornote{Equal contribution.}
\affiliation{%
  \institution{Computer Science, \\University of Illinois Urbana-Champaign}
  \city{Champaign}
  \state{Illinois}
  \country{USA}}
\email{dinasa2@illinois.edu}

\author{Mengke Wu}
\authornotemark[1]
\affiliation{%
  \institution{Information Sciences, \\University of Illinois Urbana-Champaign}
  \city{Champaign}
  \state{Illinois}
  \country{USA}}
\email{mengkew2@illinois.edu}

\author{Zijian Ding}
\affiliation{%
  \institution{College of Information, \\University of Maryland}
  \city{College Park}
  \state{Maryland}
  \country{USA}}
\email{ding@umd.edu}

\author{Jessie Chin}
\affiliation{%
  \institution{Information Sciences, \\University of Illinois Urbana-Champaign}
  \city{Champaign}
  \state{Illinois}
  \country{USA}}
\email{chin5@illinois.edu}

\renewcommand{\shortauthors}{Quan, et al.}

\begin{abstract}
Most AI systems today are designed to manage tasks and execute predefined steps. This makes them effective for process coordination but limited in their ability to engage in joint problem-solving with humans or contribute new ideas. We introduce MultiColleagues, a multi-agent conversational system that shows how AI agents can act as colleagues by conversing with each other, sharing new ideas, and actively involving users in collaborative ideation processes. In a within-subjects study with 20 participants, we compared MultiColleagues to a single-agent baseline. Results show that MultiColleagues fostered stronger perceived social presence, and participants rated their outcomes as higher in quality and novelty, with more elaboration during ideation. These findings demonstrate the potential of AI agents to move beyond process partners toward colleagues that share intent, strengthen group dynamics, and collaborate with humans to advance ideas.
\end{abstract}



\begin{teaserfigure}
\centering
\includegraphics[width=0.79\textwidth]{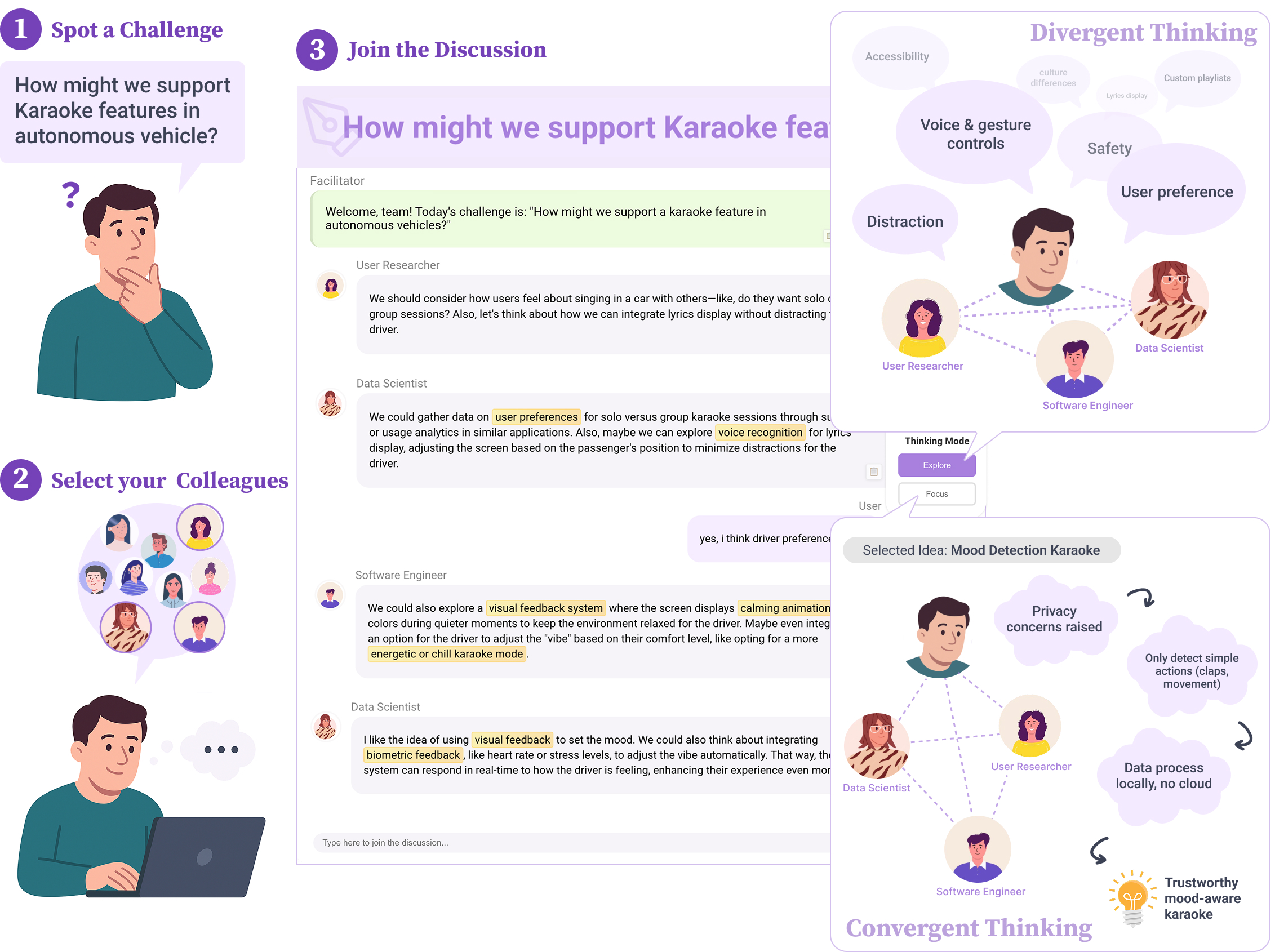}
\caption[Core scenario of MultiColleagues.]{\textbf{MultiColleagues: An Illustrative User Scenario of AI-Supported Ideation.}
P10 introduces a karaoke-in-autonomous-vehicle challenge (Step 1), selects three AI colleagues (Step 2), then moves from \textit{Explore} (divergent ideas) to \textit{Focus} (convergence) on \textit{Trustworthy mood-aware Karaoke} with privacy, simplicity, and local data processing (Step 3).}
\Description{Participant 10 scenario with MultiColleagues. The figure depicts Participant 10 identifying a challenge, selecting role-differentiated colleagues, and joining the discussion. Through explore and focus modes, divergent and convergent thinking are applied to generate and refine ideas.}
\end{teaserfigure}

\maketitle

\def \RQO {\textbf{RQ1}: ?}

\def \RQT {\textbf{RQ2}: ?}

\section{Introduction}

\label{sec:intro}

Recent adoption of large language models (LLMs) has moved from deployments as “copilot” tools toward more dynamic roles in collaborative settings. Early applications positioned LLMs as tools or judges \cite{brown2020languagemodelsfewshotlearners, ouyang2022traininglanguagemodelsfollow}, automating tasks such as summarization, programming assistance, or quality assessment that humans found repetitive or peripheral \cite{mastropaolo2023copilot, mozannar2024programming}. Although useful, these applications positioned artificial intelligence (AI) as automation rather than as a partner in collaboration \cite{hwang2022toolate, lim2009and}, whereas emerging multi-agent frameworks suggest a broader transition \cite{dubey2020haco, bansal2021team}. Systems such as AutoGen \cite{wuAutoGenEnablingNextGen2023} and CrewAI \cite{crewai2024} demonstrate how multiple LLMs can be coordinated under structured protocols, with agents adopting complementary roles such as planner, critic, or explainer. LLMs are moving beyond single-task execution to acting as participants that contribute perspectives in a team process.

The growing use of multi-agent frameworks raises a deeper question: whether LLMs can be experienced not only as tools but also as colleagues in collaborative work. Human–model collaboration is often most productive when it leverages asymmetries in capability, with people contributing judgment, values, and imagination while LLMs provide scale, recall, and breadth \cite{Dellermann_2019, amershi2019guidelines}. Moving beyond copilot metaphors requires examining whether users can experience these systems as team members with complementary strengths.

Ideation provides a representative context to probe this question. Research in creativity shows that human strengths remain decisive in the early stages of idea generation. At the same time, studies show that LLM assistance can expand the number and diversity of ideas, while also risking homogenization and over-reliance \cite{bucinca2021overreliance, lu2024llm}. Multi-agent personas present a promising approach, as they can emulate interdisciplinary team dynamics, introduce diverse perspectives, and enable humans to retain strategic oversight. Interaction paradigms play a central role in shaping this experience. Roundtable exchanges, hierarchical supervision, or progressive disclosure influence how people engage with multiple agents and how cognitive load is managed \cite{pereira2023johnny, Luger2016LikeHA}. Besides, role-playing structures task allocation and heightens social presence, which makes collaboration feel closer to working with teammates \cite{lee2019webuildai, ashktorab2019resilient}. 

Recent work has examined distinct facets of this design space (Table \ref{tab:litreview_comparison}), with systems emphasizing divergent and convergent structuring through staged debate and role play \cite{lu2024llm, coexploreds2025}, double diamond–inspired workflows \cite{supermind2024}, and  role-based orchestration that rotates perspectives to support reflection \cite{zhang2024filterbubbles, pang2025synthetic}. These approaches highlight different core elements of collaborative ideation but have largely been explored in separation. However, real-world brainstorming blends these elements as teams shift between divergent and convergent thinking, draw on multiple viewpoints, and rely on facilitation to stay aligned. This gap between fragmented system mechanisms and integrated human practice motivates systems that better reflect how collaborative ideation unfolds. 
MultiColleagues addresses this gap by integrating these mechanisms into a unified multi-agent system. Unlike workflow-based systems which structure idea generation through predefined stages \cite{supermind2024, coexploreds2025}, MultiColleagues models team cognition by having AI personas build on each other's reasoning, surface differences, and adapt through dynamic turn-taking and facilitator oversight.
Building on these approach, we introduce MultiColleagues, a multi-agent conversational system that brings together diverse AI personas for co-ideation while centering the human as facilitator-in-chief. The system follows three design goals: (1) supporting shifts between divergent and convergent thinking, (2) engaging diverse viewpoints to expand the idea space, and (3) providing clear features that preserve human oversight in collaboration.

To examine the value of this social orchestration paradigm, we conducted a within-subjects study that compares MultiColleagues with a state-of-the-art single agent system that reflects the common AI as tool model in ideation practice. This contrast provides a clear basis for assessing whether an AI as colleague design changes how users collaborate, explore ideas, and perceive the interaction. Guided by this comparison, we investigate three research questions:
\begin{enumerate}
\item \textbf{RQ 1}: How do role-taking patterns and perceptions of social presence shape the collaborative atmosphere and user engagement?

\item \textbf{RQ 2}: How does exposure to multiple AI-Colleague perspectives influence users' ideation processes and perceived outcome quality and novelty?

\item \textbf{RQ 3}: How do system design features support or constrain support during creative ideation?
\end{enumerate}

\section{Related Work}

\newcommand{\cmark}{\ding{51}}
\newcommand{\xmark}{\ding{55}}
\newcommand{\na}{\textemdash}

\definecolor{GoodGreen}{RGB}{0,128,0}
\definecolor{BadRed}{RGB}{200,0,0}
\newcommand{\yes}{\textcolor{GoodGreen}{\ding{51}}}
\newcommand{\no}{\textcolor{BadRed}{\ding{55}}}

\begin{table*}
\centering
\caption{Comparison of multi-agent systems. Icons: {\protect\yes} yes, {\protect\na} not applicable. 
\textbf{Turn} = dynamic turn selection; \textbf{Div--Conv} = divergent–convergent phases; \textbf{Orch.} = human-facing orchestration; 
\textbf{Pers.} = presence of explicit agent identities or role styles; 
\textbf{In-situ} = interactive study; \textbf{Pick} = user chooses agents; 
\textbf{Ctrl.} = controlled (within-subjects) vs. baseline, with $^{*}$ indicating comparison against ChatGPT.}
\footnotesize
\setlength{\tabcolsep}{5pt}
\begin{tabularx}{\linewidth}{l *{4}{c} *{3}{c} X}
\toprule
\multirow{2}{*}{\textbf{System}} &
\multicolumn{4}{c}{\textbf{System}} &
\multicolumn{3}{c}{\textbf{User Study}} &
\multirow{2}{*}{\textbf{Remarks}} \\
\cmidrule(lr){2-5}\cmidrule(lr){6-8}
& \textbf{Turn} & \textbf{Div--Conv} & \textbf{Orch.} & \textbf{Pers.}
& \textbf{In-situ} & \textbf{Pick} & \textbf{Ctrl.} & \\
\midrule

LLM Discussion (COLM'24) \cite{lu2024llm} &
  & \yes &  & \yes &
  \na & \na & \na &
Agents as process partners in agent–agent debate with convergence to boost model creativity on benchmarks. \\

\addlinespace[2pt]
SWTW (CHI'24) \cite{zhang2024filterbubbles} &
  &  & \yes & \yes &
  \yes &  &  &
Agents as a guidance panel for progressive exposure in media reading. \\

\addlinespace[2pt]
Weaver (CHI EA'25) \cite{pang2025synthetic} &
  &  & \yes & \yes &
  \yes & \yes & \yes$^{*}$ &
Advisory round-table with next-speaker and summaries to surface impacts. \\

\addlinespace[2pt]
Supermind (CI'24) \cite{supermind2024} &
  &  &  &  &
  \yes &  & \yes$^{*}$ &
Single agent tool using double-diamond inspired moves for guided exploration. \\

\addlinespace[2pt]
CoExploreDS (CHI'25) \cite{coexploreds2025} &
  & \yes &  &  &
  \yes &  & \yes &
Design space tool using a div-conv approach for structured exploration. \\

\specialrule{.1em}{.4em}{.4em}
\textbf{MultiColleagues (our work)} &
\yes & \yes & \yes & \yes &
\yes & \yes & \yes$^{*}$ &
\textbf{Agents as colleagues for co-ideation with Explore/Focus and human-paced facilitation. } \\

\bottomrule
\end{tabularx}
\label{tab:litreview_comparison}
\end{table*}

\subsection{From Tools to Colleagues: The Evolution of Human–AI Collaboration}

Over recent years, large language models (LLMs) have emerged as transformative tools across a wide range of applications, demonstrating state-of-the-art performance in natural language processing and knowledge-intensive tasks \cite{yang2024harnessing,lappin2024assessing}. They have been widely adopted in domains such as software development \cite{wu2024oscopilot,cui2024field_copilot}, writing and creativity support \cite{clark2022wordcraft, wu2025mbti} and scientific experimentation \cite{bran2023chemcrow}. Across these domains, LLMs have primarily functioned as assistants or evaluators, supporting human work by improving efficiency and facilitating decision-making \cite{qin2023chatgpt, yuan2022wordcraft}. Nevertheless, human-in-the-loop (HITL) involvement remains indispensable. Research has shown that relying on humans solely as “reviewers” can introduce risks, such as decision-making risks \cite{schoeffer2024dmrisk, ghosh2025yes}, reliability and transparency risks \cite{amirizaniani2024llmauditorframeworkauditinglarge}, systemic risks \cite{reid2025risk}, and ethical risks \cite{peter2025, akbulut2024all, kocaballi2022design}. Additional HITL studies across domains reinforce the irreplaceable role of humans and confirm that human involvement enhances accuracy and reliability \cite{li2025llmbasedautomatedgradinghumanintheloop, hula2024}.
Building on this foundation, human-centered and mixed-initiative frameworks argue that the future of AI lies in augmentation rather than replacement, coupling higher levels of automation with sustained human control \cite{shneiderman2022human,amershi2019guidelines}. As LLMs grow more capable, systems are gradually shifting toward proactive collaboration. Multi-agent frameworks such as AutoGen and CAMEL operationalize LLMs as differentiated collaborators \cite{li2023camel,wuAutoGenEnablingNextGen2023}, where models can assume specific roles, engage in mutual critique, and coordinate planning activities. Through such structured interactions, they approximate team-like collaboration and move toward systems where LLMs function less as tools and more as partners \cite{du2023improving, liu2023dynamic}.

\subsection{Multi-Agent LLMs and Teamwork Dynamics}
The growing capabilities of recent years’ LLMs have motivated increasing interest in multi-agent frameworks as a way to extend the scope and complexity of applications. A first major direction examines \textit{task-decomposed collaboration}, where models are assigned complementary roles such as planning \cite{wuAutoGenEnablingNextGen2023}, execution \cite{wang2024openhands}, or debugging \cite{dibia2024autogen}. Within this strand, researchers have proposed different coordination paradigms, including dialogic debate \cite{du2023improving,liang2023encouraging} and hierarchical supervision \cite{guo2024survey,wu2024oscopilot}. While these systems demonstrate gains in reasoning, factuality, and coding ability, their evaluations remain largely confined to benchmark settings \cite{hendrycks2020measuring} rather than interactive use.  

\begin{figure*}[t]
    \centering
    \includegraphics[width=\linewidth]{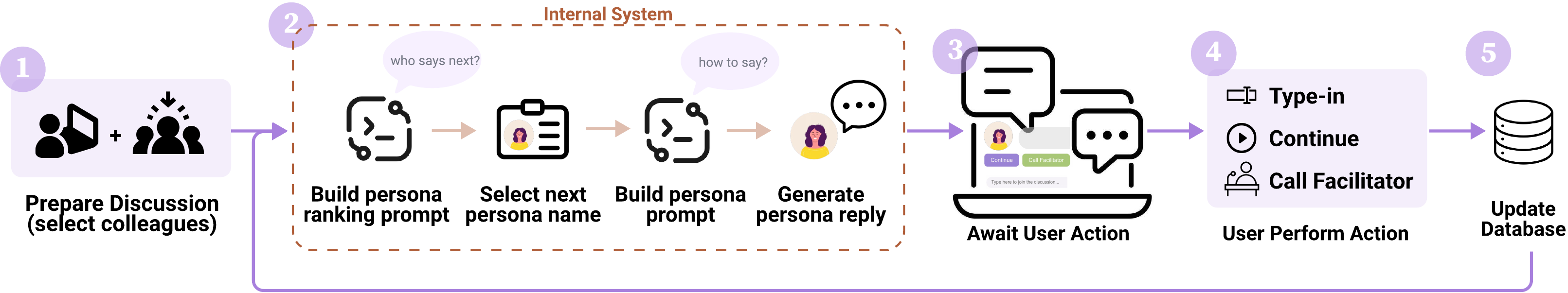}
    \caption{\textbf{System Workflow for Persona-Guided Discussions.} 
    This diagram illustrates the end-to-end workflow of the persona-guided discussion system across five steps. 
    The process begins with preparing the discussion by selecting AI colleagues (Step 1). The internal system first builds a persona ranking prompt, next selects the next persona, then builds a persona prompt and generates the corresponding reply (Step 2). The generated output is presented to the user, and the system awaits user action (Step 3). The user performs one of three actions: type a response, continue with the system-generated reply, or call a facilitator for support (Step 4). All logs are stored in the database (Step 5).}
    \Description{System workflow of persona-guided discussions. The figure shows five steps: Step 1, preparing discussion by selecting colleagues; Step 2, system ranks and selects a persona and generates a reply; Step 3, output is presented, awaiting user action; Step 4, user responds, continues, or calls a facilitator; Step 5, all logs are stored in the database.}
    \label{fig:systemflow}
\end{figure*}

A second line of work explores \textit{persona-driven role play}. Recent studies show that AI generated personas are sufficiently coherent to function as stable identities in collaborative settings \cite{schuller2024llmpersonas,smrke2025exploring,salminen2024deus,jung2025personacraft, pang2025synthetic}, enabling LLMs to convincingly simulate interdisciplinary teamwork by adopting distinct identities. This stream highlights how personas shape more human-like interaction styles \cite{park2023generative,li2023chatharuhi,wei2023multi,shao2023character}, enhance engagement \cite{wang2024rolellmbenchmarkingelicitingenhancing,cui2023thespian}, and diversify task performance through structured impersonation \cite{li2023camel}. Examples range from predefined expert roles in CAMEL \cite{li2023camel}, which bring complementary perspectives to task execution, to generative agents that exhibit emergent social behaviors in daily scenarios \cite{park2023generative,shanahan2023role}. Collectively, this work points to the potential of role differentiation for strengthening social presence and aligning collaboration with human expectations.  

A third line of research focuses on \textit{coordination mechanisms} that sustain coherence across longer or more complex interactions. Efforts here include shared memory, scheduling, and blended model outputs. Skeleton-guided reasoning \cite{ning2023skeleton}, model fusion \cite{jiang2023llm}, and collaborative decoding strategies \cite{sun2023corex} exemplify how multiple agents can combine strengths to tackle tasks beyond the capacity of a single model. At the same time, across all three directions, prior work consistently emphasizes the importance of human oversight for aligning decisions with user intent and preventing failures under high autonomy \cite{bansal2021team,amershi2014rolehuman,christiano2017deep}.

As shown in Table~\ref{tab:litreview_comparison}, recent work has begun to explore multi-agent systems in interactive settings. LLM Discussion adopts a three-phase agent–agent debate with role-play to enhance originality and elaboration of model outputs \cite{lu2024llm}. SWTW introduces progressively contrasting roles in media reading, using orchestration and gamified puzzles to mitigate filter bubbles and deepen reflection \cite{zhang2024filterbubbles}. Weaver organizes advisory-style roundtables with speaker rotation and summaries to anticipate broader social impacts  \cite{pang2025synthetic}. While each contributes valuable orchestration strategies, they do not combine dynamic turn-taking, divergent–convergent phases, and user-facing facilitation within controlled, in-situ studies. Motivated by these gaps, MultiColleagues advances co-ideation by combining dynamic turn-taking, explicit divergent–convergent shifts, and interactive orchestration, enabling a controlled study of how role differentiation and multi-agent dynamics shape collaborative ideation.

\subsection{GenAI-Assisted Ideation}

Recent works on AI-assisted ideation spans a wide range of approaches that enrich creative exploration. Early systems emphasized retrieving relevant concepts to spark new ideas, either through retrieval-based pipelines that surfaced related crowd-generated content \cite{ideahound2016} or through similarity-driven semantic mapping that highlighted adjacent concepts \cite{macneilProbMapAutomaticallyConstructing2021, xuIdeateRelateExamplesGallery2021}. With the rise of generative models, newer tools embed LLM outputs into interactive structures that support higher-level reasoning, such as template-guided reflection \cite{xuJamplateExploringLLMEnhanced2024}, analogical inspiration cards derived from LLM-generated concepts \cite{kangBioSparkAnalogicalInspiration2025}, recombined scientific facets for idea expansion \cite{radensky2025scideatorhumanllmscientificidea}, and causal diagrams that help users explore conceptual relationships \cite{huang2023causalmapper}. Other systems introduce varied perspectives by incorporating conversational prompting \cite{rayanExploringPotentialGenerative2024} or using role conditioning to simulate expert viewpoints \cite{liu2025personaflow}. Collectively, these works demonstrate the breadth of mechanisms through which generative AI can support creative thinking.

With recent advances in large language models, conversational systems have become effective day to day ideation tools. ChatGPT supports fast concept expansion, iterative refinement, and low effort interaction, making it a widely used single agent workflow for brainstorming and creative exploration \cite{openai2024chatgpt}. Reflecting its central role in practice, recent creativity-support systems often build on GPT or benchmark their structured interfaces against ChatGPT-style prompting \cite{wang2025aideation, supermind2024, shen2025ideationweb}.
Despite this widespread adoption, single-agent LLMs remain constrained by their autoregressive nature, which tends toward convergent trajectories \cite{Wadinambiarachchi2024fixation}, and by the limited viewpoint available from a single generative source. These constraints have motivated interest in introducing additional structure or complementary perspectives into the ideation process. Existing work explores individual mechanisms such as divergence–convergence structuring \cite{lu2024llm, coexploreds2025}, Double Diamond workflows \cite{supermind2024}, or modeled perspective shifts \cite{zhang2024filterbubbles, pang2025synthetic}. Multi-agent approaches have also emerged as a way to distribute reasoning across personas to encourage more diverse exploration \cite{fukumura2025creativity, ghosh2025yes}. These mechanisms support different parts of the ideation process but remain largely separate, without enabling the coordinated interaction of perspectives and phases typical of real brainstorming. MultiColleagues integrates these elements within a single system, which offers a unified, team-based alternative to a single-agent ChatGPT workflow.

\section{System Design}

\subsection{Design Goals and Implementation}
Our reflections on prior work in conversational agents for human–AI collaboration led us to articulate the following design goals for our MultiColleagues system:

\textbf{(1) DG1: Support adaptive Human AI co-ideation dynamics.}
We focused on how co-ideation tools can help users shift smoothly between expansive exploration and targeted evaluation while maintaining control over the creative trajectory. Creative problem solving is understood as an iterative cycle of divergence and convergence that benefits from structured scaffolds \cite{designcouncil2005doublediamond}. Recent systems such as Supermind Ideator show that reflective moves and guided reframing can reduce fixation and improve idea quality \cite{supermind2024}. Building on this work, our system supports flexible transitions across ideation phases as users’ goals evolve.

\textbf{(2) DG2: Enable rich, multi-perspective co-ideation.}
Work on boundary objects demonstrates that shared artifacts help people coordinate across disciplinary differences by making reasoning and assumptions available for joint inspection \cite{star1989institutional, halpern2013designing}. This aligns with creativity-support research showing that exposure to varied perspectives expands the idea space and reduces the risk of narrow, single-voice trajectories \cite{crowdboard2017, ideahound2016}. More recent work with role-based and persona-based AI further shows that structured viewpoints can reveal tensions, question defaults, and surface assumptions that remain hidden when users interact with a single assistant \cite{pang2025synthetic, zhang2024filterbubbles, ghosh2025yes, liu2025personaflow}. Building on this foundation, our system focuses on providing accessible and differentiated viewpoints that function as lightweight boundary objects, supporting interdisciplinary-style exploration even when users work alone and helping them avoid constrained or homogeneous idea paths.

\textbf{(3) DG3: Facilitate purposeful and transparent Human-AI collaborative control.}
Human–AI teaming studies highlight that effective collaboration depends on shared awareness and clear communication that help calibrate trust between human and machine partners \cite{zhang2021ideal, zhang2023investigating}. In creative and co-productive settings, coordination structures such as visible turn-taking, recognizable contribution types, and explicit interaction points help users interpret system actions and maintain oversight of the collaborative process \cite{winston2017turn, guzdial2019turnbased, kantosalo2020modalities}. Transparency and user control over automation are also linked to stronger engagement and more intentional steering of AI-generated content \cite{amershi2019guidelines, zhang2025exploring}. Guided by this literature, our system emphasizes interaction patterns that make system behavior legible and predictable, offering clear interaction signals and visual scaffolds that allow users to work with AI colleagues while maintaining strategic control.

To address our design goals, we designed and developed \textbf{MultiColleagues}, a human–AI collaborative conversational platform where multiple AI personas participate in structured brainstorming alongside the user. The system architecture follows a two-tier design pattern: presentation layer (React frontend) and application layer (Flask API). The system integrates OpenAI Large Language Model \textbf{GPT-4o} for natural language generation (see Figure \ref{fig:systemflow}).

\subsubsection{Adaptive Thinking Transition Mechanisms}
\begin{figure*}[t]
\centering
\includegraphics[width=\linewidth]{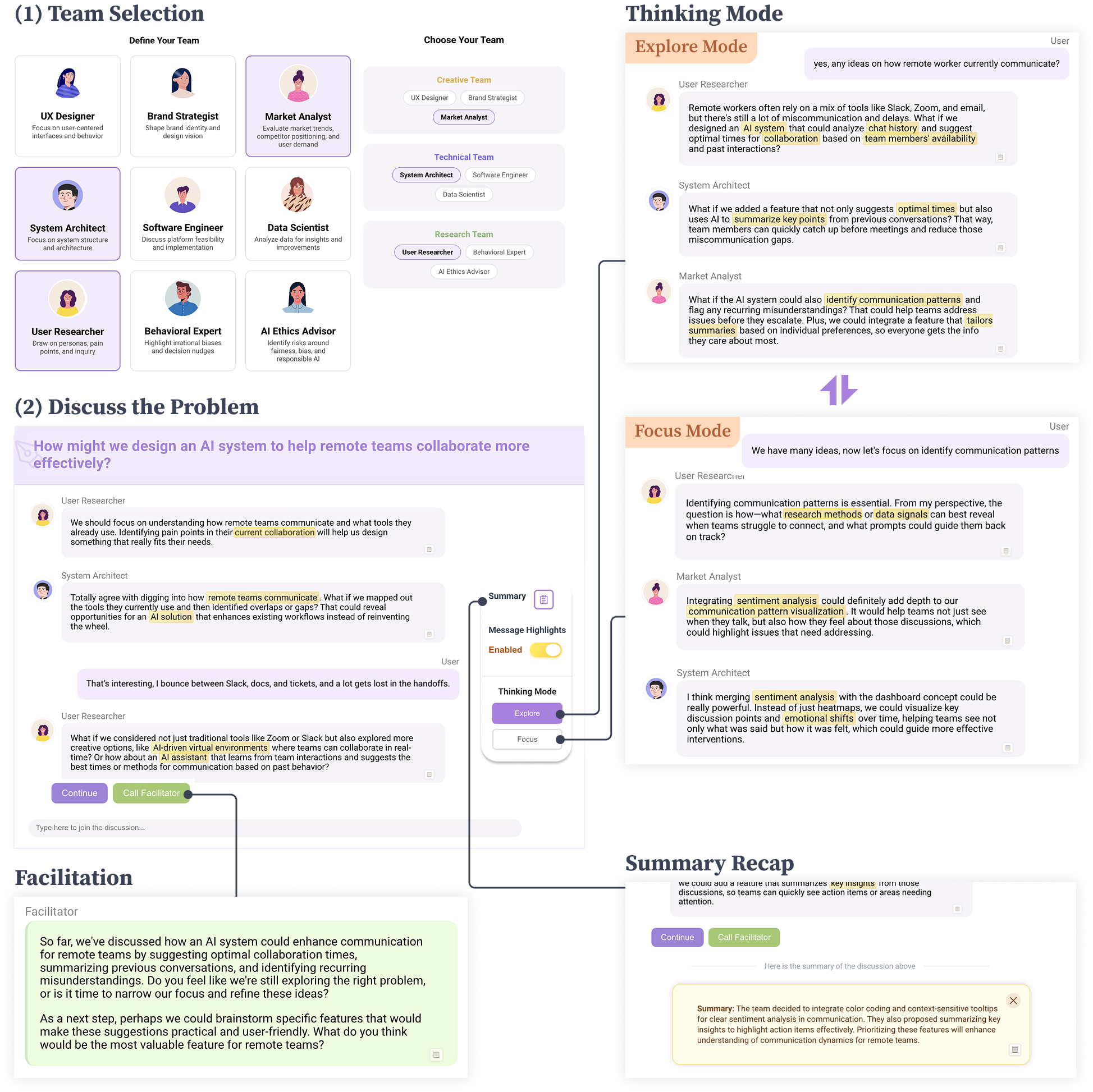}
  \caption[Usage flow of the Multi-agent Conversational System for co-ideation]{\textbf{Usage flow of the Multi-agent Conversational System for co-ideation.}
  (1) Users first \textit{select a team} of persona experts.
  (2) They then \textit{discuss the problem}, with personas contributing from their own perspectives.
  The system supports two \textit{thinking modes}: \textbf{Explore mode} expands ideas broadly, while \textbf{Focus mode} refines and synthesizes them.
  A \textit{summary recap} highlights key points, and \textit{facilitation} is triggered either by user request or automatically by the system when guidance is needed.}
  \Description{Composite UI showing team selection, problem discussion, Explore and Focus modes, summary recap panel, and facilitation panel.}
  
  \label{fig:completeflow}
\end{figure*}

To support adaptive thinking transitions in collaborative ideation (DG1), we implemented a \textit{dual-mode switching framework} grounded in the double diamond design methodology, explore mode and focus mode, as illustrated within the usage flow (Fig.~\ref{fig:completeflow}). The Explore mode emphasizes breadth and diversity of viewpoints. During this mode, different colleague experts are encouraged to expand the space of possibilities, each approaching the problem from their own perspectives and generating ideas freely without immediate concern for constraints. Focus mode emphasizes depth, clarity, and actionable outcomes. In this stage, colleagues shift toward evaluating, filtering, and aligning ideas, working from their own roles to concentrate on feasibility and actionable outcomes.

Our system provides effective ideation through systematic alternation between divergent exploration phases (expanding problem and solution spaces) and convergent synthesis phases (evaluating, refining, and consolidating ideas). This template-based approach ensures that thinking mode transitions extend beyond interface changes to actually modify AI thinking processes in alignment with human creative phases. 

\textit{Strategic interaction points} are integrated after each AI response to preserve human agency and prevent cognitive saturation, as we found that uninterrupted AI generation creates information overload and diminishes human creative contribution and strategic oversight capacity \cite{Abkemeier2020CognitiveLT}. These pause points enable participants to make explicit choices to ``Continue'' autonomous AI discussion or ``Call Facilitator'' for guided intervention, which preserves human cognitive control by preventing information saturation and enabling reflective engagement with AI-generated content before proceeding to subsequent ideation phases (DG3).

\textit{An AI Facilitator} functions as a meta-cognitive regulator that monitors conversation dynamics and intervenes at calculated intervals when discussions deviate from productive ideation patterns, lack adequate synthesis, or fail to incorporate diverse participant perspectives \cite{Flavell1979MetacognitionAC}. The facilitator employs real-time conversation analysis to identify critical transition points between diamond phases. It explicitly prompts users with a quick overview of the current conversation history, and to reflect on whether the team should continue divergent exploration or transition toward convergent evaluation, and provides structured progress synthesis to prevent cognitive fragmentation across extended ideation sessions.

\subsubsection{Multi-Persona Orchestration System}
To enable rich, multi-perspective co-ideation (DG2), the system implements a persona orchestration framework that instantiates a diverse roster of AI colleagues with distinct \textit{professional backgrounds}, \textit{communication styles}, and \textit{domain expertise} (see Figure \ref{fig:orchestration} in Appendix \ref{appendix:d}.1). Each persona is constructed through structured configurations defining behavioral instructions for communication patterns, specialized knowledge domains that establish topical authority, and participation patterns that govern engagement frequency. These persona templates are drawn from prior works on multi-agents' interdisciplinary collaboration \cite{hong2024metagptmetaprogrammingmultiagent}, which show that encoding standardized workflows into multi-agent prompts improves coordination and reduces cascading errors. To support purposeful and transparent collaborative control, we implemented a structured turn-taking mechanism that manages how personas contribute during ideation (DG3). Personas introduce their perspectives in stages to reduce cognitive load and maintain contribution quality. Each persona first produces an individualized initial response, after which an AI-driven evaluation selects the opening speaker based on relevance and conversational coherence.
For subsequent turns, a dynamic ranking mechanism identifies the persona most suited to speak next by assessing contextual relevance, conversation history, and unexpressed perspectives, with a small randomization factor (20\%) to avoid rigid patterns. Once a persona is selected, the system retrieves its prompt template and combines it with the full conversation context, the user's most recent input comments, and the current orchestration state (focus vs. explore). Guided by this structured input, the language model generates the persona’s output. To sustain longer dialogues, a conversational history compression pipeline is applied when the message count exceeds a threshold. As shown in Figure \ref{fig:compact}, recent turns’ history is kept in full while older persona contributions are summarized, allowing the system to maintain immediate context while compactly representing earlier perspectives. This compression design ensures efficiency and coherence in multi-colleague conversations (see Appendix \ref{appendix:d}.2). Detailed prompt templates for persona creation, first-speaker selection, persona ranking, and response generation are provided in Appendix \ref{appendix:c}.

\begin{table*}[t]
\centering
\caption{Participant demographics and creativity scores.}
\small
\begin{minipage}{0.78\linewidth}
\centering
\setlength{\tabcolsep}{5pt}
\begin{tabular}{@{} l c p{2.8cm} p{4.0cm} r @{}}
\toprule
\textbf{ID} & \textbf{Age} & \textbf{Occupation} & \textbf{Specialization} & \textbf{Creativity} \\
\midrule
P1  & 25-29 & Master Student         & Communication                          & 2.09 \\
P2  & 25-29 & PhD Student            & Material Science \& Engineering        & 4.91 \\
P3  & 25-29 & Professional           & Computer Science                       & 4.91 \\
P4  & 20-24 & Master Student         & AI Research                            & 5.18 \\
P5  & 25-29 & Professional           & Information Science                    & 7.00 \\
P6  & 25-29 & PhD Student            & Information Science                    & 6.27 \\
P7  & 20-24 & Master Student         & NLP                                    & 4.73 \\
P8  & 25-29 & Professional           & UX Research                            & 4.64 \\
P9  & 25-29 & PhD Student            & Computer Science                       & 6.55 \\
P10 & 25-29 & Professional           & Design                                 & 5.27 \\
P11 & 20-24 & PhD Student            & Computer Science                       & 4.64 \\
P12 & 30-34 & PhD Student            & Voice Interaction                      & 6.27 \\
P13 & 20-24 & Undergraduate Student  & HCI                                    & 5.36 \\
P14 & 25-29 & PhD Student            & Virtual Reality                        & 6.09 \\
P15 & 35-39 & PhD Student            & AI Research                            & 5.27 \\
P16 & 30-34 & PhD Student            & HCI                                    & 6.45 \\
P17 & 20-24 & Master Student         & Virtual Reality                        & 6.00 \\
P18 & 25-29 & Professional           & Design                                 & 6.00 \\
P19 & 25-29 & PhD Student            & Machine Learning                       & 5.18 \\
P20 & 30-34 & PhD Student            & Information Science                    & 6.55 \\
\bottomrule
\end{tabular}
\label{tab:participants}
\end{minipage}
\end{table*}

\subsubsection{User-Friendly Interface and Interaction Design}
To maintain strategic oversight in multi-agent ideation while preventing passive consumption of AI output, we implement user-friendly interfaces and interaction mechanisms that enable clear control and integration points throughout collaborative discussions (DG3). Beyond the conversation flow controls described in DG1, the system provides clear visual cues through distinctive persona profile pictures and role-based message styling that enable users to quickly identify different AI perspectives and track individual contributions across extended discussions. An intelligent highlighting system with user-controlled visibility allows participants to manage information density by toggling keyword emphasis on or off, which supports cognitive load management and preserves access to AI-generated insights. The interface enforces each persona's conversation wording limits and structured message threading that enhances readability during multi-agent exchanges. Besides, the thinking mode controls enable users to explicitly switch between ``explore'' mode (divergent thinking) and ``focus'' mode (convergent thinking), with clear visual indicators showing the current cognitive state and immediate effects on subsequent AI behavior.

\subsection{Usage Scenario} 
To illustrate how participants engage with the multi-agent co-ideation system, we present a typical user journey (see Figure \ref{fig:completeflow}). The participant begins by selecting three AI personas (User Researcher, System Architect, and Market Analyst) and submitting the problem statement \textit{``How might we design an AI system to help remote teams collaborate more effectively?''} The system generates initial thoughts from each persona and presents the User Researcher as the first speaker, who raises questions about user pain points in remote collaboration. During the initial exploration phase, the participant primarily operates in ``explore'' mode, alternating between clicking ``Continue'' to observe autonomous AI discussions and actively joining the conversation by submitting their own insights. After approximately six AI responses covering topics ranging from technical infrastructure concerns to user experience considerations, the participant clicks ``Call Facilitator'' to request guidance on discussion direction. The AI facilitator provides a synthesis of perspectives shared so far and prompts the participant to consider whether the team should continue exploring the problem space or begin focusing on specific solution approaches. At this point, the participant switches to ``focus'' mode and submits a message, narrowing the scope to ``real-time collaboration tools for creative teams.'' The AI personas now operating in convergent thinking mode, begin evaluating and synthesizing the discussed ideas, with the System Architect proposing specific technical architectures while the User Researcher focuses on user-related principles. The participant carefully reviews the intelligent highlighting phrases to identify key concepts to consolidate the emerging solution framework.

\section{Study Design}
To examine how different AI interaction paradigms influence creative collaboration, we designed our study structured as follows: participant recruitment and characteristics (Section \ref{participant}), detailed procedure including task design (Section \ref{procedure}), and data collection and analysis methodology (Section \ref{evaluation}).

\subsection{Participants} \label{participant}
As shown in Table~\ref{tab:participants}, our study recruited 20 participants through university mailing lists and professional networks (9 males, 11 females), aged 20-39 years (M = 26.7, SD = 3.8). 15 participants were undergraduate to PhD students pursuing degrees in fields such as computer science, information science, HCI, AI research, and communication, while the remaining 5 were early-career professionals working in technology, research, and design-related roles. All participants met our eligibility criteria, including relevant domain backgrounds and prior experience with creative problem-solving or AI-driven tools. The study received the university’s IRB approval, and each participant was compensated \$20.
Table~\ref{tab:participants} summarizes participants’ demographics along with their creativity scores (M = 5.39, SD = 1.16). Scores were calculated as the average of 11 items from a 7-point Likert-scale creativity assessment (see Appendix~\ref{appendix:a}.1).

\subsection{Study Procedure} \label{procedure}
Each study session lasted approximately 70 minutes and was conducted remotely via Zoom with a researcher present to observe interactions and provide technical support when needed. We used a within-subjects design to reduce individual variability in ideation skill and cognitive style, allowing direct comparison of collaboration patterns across conditions. All participants interacted with both \textbf{MultiColleagues} and \textbf{ChatGPT}, which shared the same underlying model \textbf{GPT-4o}, in counterbalanced order to mitigate order effects \cite{pollatsek1995use}. ChatGPT was selected as the baseline as it represents the dominant single agent, chat based workflow for everyday ideation \cite{supermind2024, shen2025ideationweb, wang2025aideation}. This choice increases ecological validity by grounding the comparison in the AI as tool paradigm that users already rely on \cite{sinlapanuntakul2025impacts}. While an ablated multi agent baseline could isolate specific mechanisms, our goal was not to test individual features in isolation but to examine whether a socially orchestrated AI as colleague system shifts how people brainstorm relative to this prevailing tool based model. The study procedure comprised the following phases (see Figure~\ref{fig:procedure} for a visual overview):

\begin{figure*}[h]
    \centering
    \includegraphics[width=1\linewidth]{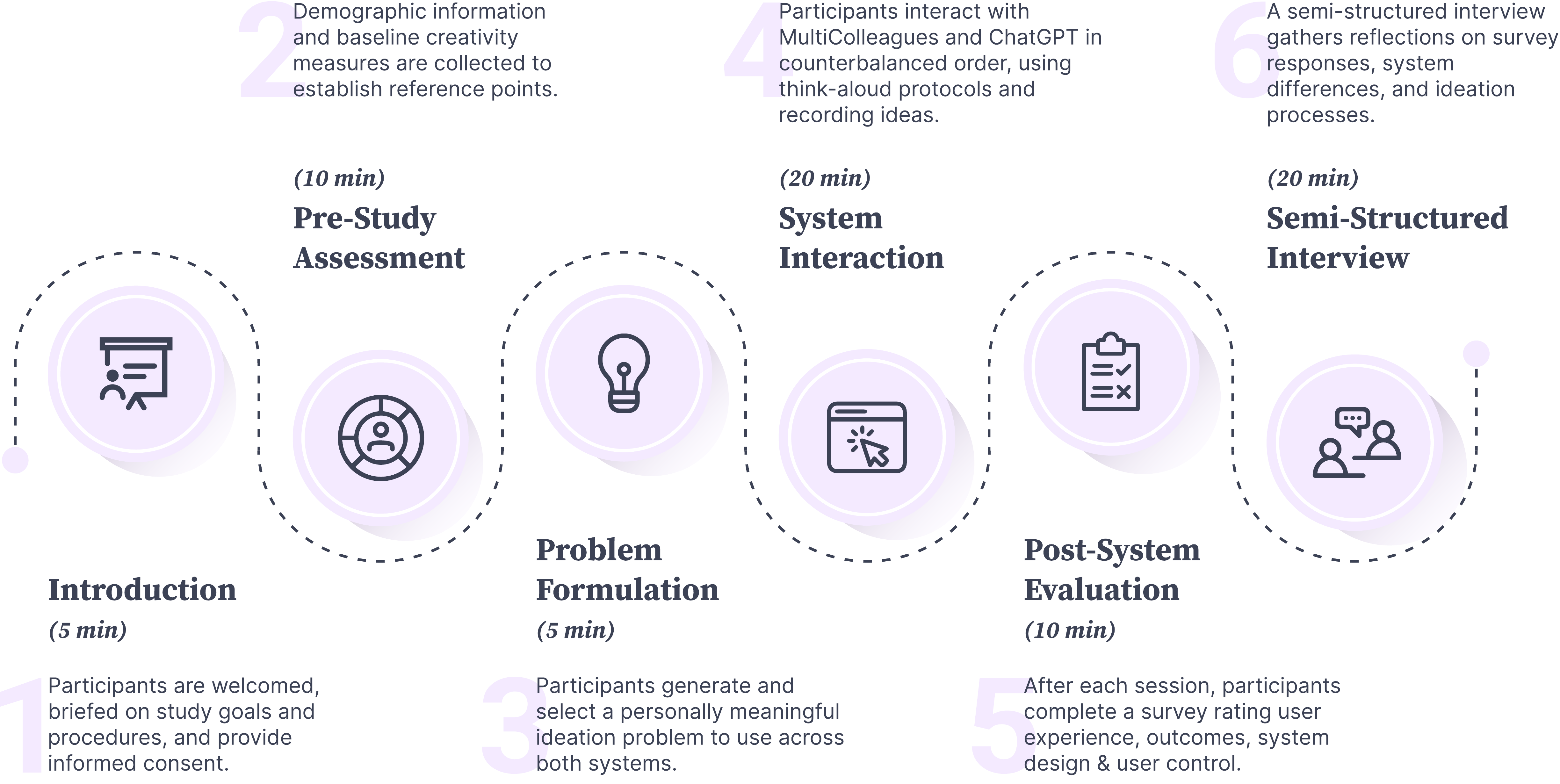}
    \caption{User Study Workflow.}
    \Description{The figure outlines six sequential stages: introduction, pre-study assessment, problem formulation, system interaction, post-system evaluation, and semi-structured interview, which shows how participants progress through the user study.}
    \label{fig:procedure}
\end{figure*}

\textbf{Introduction and informed consent} (5 minutes): Participants were briefed on study objectives, procedures, and data handling protocols before providing written informed consent.

\textbf{Pre-study assessment} (10 minutes): Participants completed demographic questionnaires capturing age, gender, educational background, and prior experience with AI-assisted creativity tools. They also completed established creativity assessment instruments adapted from validated scales \cite{runco2001ideational, tierney2002creative} to establish baseline creative capabilities (items provided in Appendix \ref{appendix:a}.2).

\textbf{Problem formulation} (5 minutes): Before experiencing systems, participants received a brief prompt on what constitutes a suitable ideation problem and chose a topic they were familiar with or personally interested in. Previous works revealed that predefined problem themes reduced participants' engagement \cite{Mohanani_2021}, therefore in the main study, participants generated their own topics of interest. The same problem was used with both systems, with a counterbalanced order to minimize order effects \cite{pollatsek1995use}.

\textbf{System interaction phase} (20 minutes): Participants were randomly assigned to one of two condition orders, with equal distribution across sequences to control order effects. Think-aloud protocols \cite{van1994think} were employed throughout both interactions to capture real-time cognitive processes. For both conditions, a split-screen setup was provided with a Google Doc on the right side for recording generated ideas following each interaction. Each ideation session was set to approximately 10 minutes to emulate fast paced brainstorming, a duration informed by our pilot study which showed that the facilitator kept discussions on track and supported more focused topic oriented exchanges. This 10-min structure also aligns with prior work using short creative sprints for AI-assisted co-ideation \cite{pang2025synthetic, qin2025timingmatters}. Our goal was to examine immediate coordination and convergence processes rather than long-term team formation. Following recent studies that allowed flexible engagement durations \cite{pang2025synthetic}, participants could extend their sessions if additional time was needed or conclude early upon idea exhaustion or fatigue.

\begin{itemize}
\item \textbf{MultiColleagues condition}: Participants received a structured tutorial (approximately 2 minutes) covering interface navigation, persona selection mechanisms, thinking mode transitions, and facilitator function utilization. Following orientation, participants engaged in a 10-minute guided brainstorming session employing the multi-agent system architecture with their self-selected problem.

\item \textbf{Baseline condition}: Participants unfamiliar with ChatGPT received a brief interface orientation focusing on conversation initiation and prompt formulation strategies. To ensure model consistency, all participants accessed \textbf{ChatGPT through the web interface using the GPT-4o model}. Participants then conducted unrestricted text-based ideation sessions using identical problem parameters.

\end{itemize}

\textbf{Post-system evaluation} (10 minutes total; 5 minutes per system): Following each system interaction, participants completed a 12-item survey on 7-point Likert scales (1 = strongly disagree, 7 = strongly agree). The instrument was adapted from established questionnaires \cite{brooke1996sus, berretta2023job, carroll2009csi, hart2006nasa, ttf1995}. Items were selected from an initial pool of 30 and refined through iterative review for clarity and relevance to our research questions; a small pilot (N = 5) ensured comprehensibility and removed redundancy. The survey evaluated three key dimensions:  \textit{Experience} (3 items), \textit{Outcomes} (3 items), and \textit{System Design \& User Control} (4 items). The complete set of 12 items is provided in Appendix~A.

\textbf{Semi-structured comparative interview} (20 minutes): The concluding interview phase employed a structured protocol designed to elicit detailed comparative reflections on participants' experiences across both systems. Participants first reviewed and elaborated on their quantitative survey responses, providing contextual explanations for their ratings. Subsequently, three targeted follow-up questions were explored: (1) the extent to which different personas provided distinctive perspectives and fulfilled unique collaborative roles, (2) how participants' perceived relationship and interaction patterns with AI agents differed between single-agent and multi-agent configurations, and (3) whether exposure to multiple personas influenced participants' ideation structuring processes or evaluative criteria for generated concepts.

\subsection{Evaluation Methods} \label{evaluation}
To answer our research questions, we adopted a mixed-methods approach integrating behavioral and perceptual measures. Quantitative data were collected from conversation histories and system interaction logs, complemented by structured post-system evaluations of usability and creativity. Qualitative data came from semi-structured interviews, capturing participants’ experiences with collaboration strategies, the outcomes of navigating different interaction paradigms, and the utility of multi-agent versus single-agent approaches. The following sections detail our evaluation methods.

\subsubsection{Comparative Analysis of User Ratings} To statistically evaluate the 12 survey data, we employed a structured analytical approach that began with thematically grouping the questions according to our research questions, followed by a non-parametric comparison of the two systems. The 12 questions were first organized into three core dimensions: (1) \textbf{Experience (RQ1)}, which captured the subjective quality of the collaboration and user engagement, as well as factors shaping those outcomes; (2) \textbf{Ideation Process and Perceived Outcomes (RQ2)}, which captured participants’ self-reported ideation process, their perceived outcome quality and novelty; and (3) \textbf{System Design \& Control (RQ3)}, which evaluated the perceived sense of control and flexibility. 
To streamline the analysis, closely related survey questions were merged into broader, more robust metrics, such as \textit{Outcome Quality \& Novelty}. Given the ordinal nature of the Likert scale data and the within-subjects design of the study, we used the non-parametric Wilcoxon signed-rank test to compare the paired ratings of MultiColleagues and Baseline for each metric. Effect sizes were calculated as the magnitude of observed differences between conditions.

\subsubsection{Thematic Analysis} Two researchers conducted a thematic analysis \cite{braun2024thematic} of interview transcripts, independently coding the data before reconciling differences and refining the codebook. Themes were organized around the study’s research questions on collaborative experience, creative outcomes, and system support. Sub-themes (e.g., “Traceable Perspectives”) emerged through iterative discussion and were grouped into role- or condition-specific insights.. 

\subsubsection{Analysis of User Idea Contribution} 

We examined the semantic structure of participants input during the ideation discussion to understand how they engaged with the task and how their ideas developed. GPT-5 with role based prompts was used to segment each user contribution into two levels. Main topics capture the central themes guiding the users reasoning. Sub topics capture the concrete points, reasoning paths, and directions expressed within those themes. We chose GPT-5 for its state-of-the-art reasoning and judgment capabilities across domains to ensure reliable
idea assessment \cite{wang2025capabilities}.
To reduce stochastic variation, each conversation was processed three times and the outputs were averaged. Two researchers then reviewed the extracted topics for alignment with the coding criteria, yielding strong inter rater reliability (Cohen's $\kappa = 0.86$) \cite{cohen1960coefficient}. Based on these annotations, we computed a branching ratio defined as the number of sub topics divided by the number of main topics, which characterizes whether a participant developed ideas in a linear form with fewer branches or in a more exploratory form with multiple parallel directions.
To further assess how participants expressed detailed concepts within their broader ideation problem, we applied a natural language toolkit for noun extraction in each user contribution \cite{bird2009natural}. Nouns offer a reasonable proxy for concept introduction and have been used to estimate conceptual density in discourse analysis \cite{kintsch1998comprehension}.

\section{Results}
\label{sec:results}

\begin{table*}[ht]
\centering
\caption{Statistical Comparison for Process \& Experience Metrics between \textbf{MultiColleagues (MC)} and \textbf{Baseline}. Results show significantly higher ratings for MC across all 3 metrics.}
\begin{tabular}{lcccccc}
\toprule
\textbf{Metric} & \textbf{Q\#} & \textbf{MC (M ± SD)} & \textbf{Baseline (M ± SD)} & \textbf{W} & \textbf{p-value} & \textbf{Effect Size ($r$)} \\
\midrule
Teammate-like Feel & Q6 & 5.75 ± 1.02 & 5.05 ± 1.15 & 17.5 & \textbf{.046*} & 0.49 \\
Complementary Strengths & Q8 & 6.05 ± 1.00 & 5.05 ± 1.64 & 2.5 & \textbf{<.01**} & 0.71 \\
Engagement \& Flow & Q10 & 5.70 ± 1.38 & 4.45 ± 1.57 & 30.0 & \textbf{.014*} & 0.63 \\

\bottomrule
\end{tabular}
\label{tab:rq2_experience}
\end{table*}

We organize the results into two parts. First, we report log analyses that establish overall interaction statistics in the MultiColleagues condition (Section 5.1). Our descriptive results capture how participants engaged with the system in terms of colleague selection patterns and message distributions, providing context for interpreting subsequent findings. The remainder of the results sections are structured around our three research questions, combining survey responses, conversation logs, and interviews. For RQ1 (Experience), we examine how role-taking patterns and perceptions of social presence shaped participants’ collaborative atmosphere and engagement. For RQ2 (Outcomes), we assess how exposure to multiple AI-Colleague perspectives influenced the quality, novelty, and organization of creative outputs. For RQ3 (System Support), we evaluate how system design features affected participants’ sense of agency, control, and flexibility during ideation.

\subsection{Overview of Participants' Interaction Statistics}
To contextualize participants’ engagement with MultiColleagues, we first examined colleague selection patterns, followed by interaction statistics across sessions. The analysis revealed substantial variance in AI colleague choices, with 95\% unique persona combinations across 20 participants. This diversity reflects strong individual differences in how participants valued expertise for ideation, suggesting that effective collaboration benefits from accommodating varied teaming preferences. 

Turning to conversational dynamics, multi-chat sessions included an average of M = 4.00 personas (SD = 1.00), ranging from three to seven colleagues. On average, each session contained 31.3 utterances, while participants contributed M = 8.3 utterances (26.7\%). To avoid overweighting the AI presence by summing across multiple colleagues, we examined the average contribution per AI colleague. Individual colleagues produced M = 5.04 utterances (SD = 1.63). This indicates that while AI colleagues collectively sustained much of the conversational flow, each colleague’s contribution was comparable in magnitude to that of the human participant, suggesting a more balanced distribution of interaction.

\begin{table*}[ht]
\centering
\caption{Statistical Comparison of Linguistic Cohesion and Pragmatic Style Metrics between \textbf{MultiColleagues (MC)} and \textbf{Baseline}. Results show a significant difference in \textit{Directness} of conversational style.}
\begin{tabular}{lccccc}
\toprule
\textbf{Metric}  & \textbf{MC (M ± SD)} & \textbf{Baseline (M ± SD)} & \textbf{W} & \textbf{p-value} & \textbf{Effect Size ($r$)} \\
\midrule
\multicolumn{6}{c}{\textbf{Linguistic Cohesion Metrics}} \\
\midrule
Narrativity & 24.40 ± 14.50 & 18.62 ± 15.62 & 59.0 & .090 & 1.58 \\
Syntactic Simplicity  & 28.62 ± 15.72 & 28.83 ± 19.62 & 92.0 & .648 & 2.46 \\
Word Concreteness & 15.13 ± 17.71 & 19.13 ± 19.24 & 82.0 & .409 & 2.19 \\
Referential Cohesion  & 27.25 ± 18.55 & 25.42 ± 20.63 & 96.0 & .756 & 2.57 \\
Lexical Diversity (MTLD) & 82.66 ± 45.02 & 71.70 ± 56.98 & 78.0 & .494 & 0.23 \\
\midrule
\multicolumn{6}{c}{\textbf{Pragmatic / Interaction Style Metrics}} \\
\midrule
Sentiment     & 4.44 ± 0.41 & 4.46 ± 0.33 & 104.0 & .985 & 2.78 \\
Formality     & 4.00 ± 0.42 & 4.08 ± 0.63 & 80.0  & .368 & 2.14 \\
Directness    & 4.76 ± 0.23 & 5.05 ± 0.44 & 30.0  & \textbf{.009**} & 0.80 \\
Relationship  & 4.39 ± 0.34 & 4.17 ± 0.44 & 88.0  & .545 & 2.35 \\
Participation & 4.29 ± 0.67 & 4.09 ± 1.07 & 87.0  & .521 & 2.33 \\
\bottomrule
\end{tabular}
\label{tab:metrics_comparison}
\end{table*}

\begin{table*}[ht]
\centering
\caption{Statistical Comparison of User Interaction Metrics between \textbf{MultiColleagues (MC)} and \textbf{Baseline}.}
\begin{tabular}{lccccc}
\toprule
\textbf{Metric} & \textbf{MC (M ± SD)} & \textbf{Baseline (M ± SD)} & \textbf{W} & \textbf{p-value} & \textbf{Effect Size ($r$)} \\
\midrule
\# of Utterances & 8.35 ± 5.79 & 4.10 ± 2.45 & 9.5 & \textbf{.001**} & 0.71 \\
Total User Words & 104.70 ± 55.85 & 51.30 ± 42.47 & 18.0 & \textbf{<.001***} & 0.73 \\
\# of Utterances per Minute & 0.65 ± 0.38 & 0.42 ± 0.23 & 125.0 & \textbf{.044*} & 0.32 \\
Total User Words per Minute & 8.12 ± 4.25 & 5.12 ± 3.67 & 119.0 & \textbf{.029*} & 0.35 \\
Average Word Count per Message & 13.47 ± 4.65 & 12.83 ± 9.12 & 68.0 & .177 & 0.31 \\
Session Duration (minutes) & 12.90 ± 2.80 & 9.80 ± 2.30 & 33.0 & \textbf{.006**} & 0.60 \\
\bottomrule
\end{tabular}
\label{tab:user_interaction}
\end{table*}

\subsection{RQ1: Collaborative Experience with Multi-AI Colleagues}
To address RQ1, we operationalized collaborative experience into three dimensions: complementary strengths, teammate-like relationships, and engagement. We observed significant differences across all, with MultiColleagues rated more positively than Baseline. We present results for each dimension, drawing on both post-system quantitative survey evaluations  (Table~\ref{tab:rq2_experience}) and interviews.

\subsubsection{Distributed Collaboration Enhances Complementary Strengths.} 
Survey results and interview reflections consistently emphasized that MultiColleagues enabled a stronger sense of distributed collaboration, where different AI colleagues contributed complementary strengths to the discussion from distinct angles. During the interview, we first invited participants to reflect on their collaborations whether they felt colleagues served distinctive roles and offered perspectives from different angles on a 7-point Likert scale. Findings confirm that most participants perceived colleagues as differentiated contributors (M = 5.85, SD = 1.09). Moreover, survey comparisons showed that participants rated the MultiColleagues condition (M = 6.05, SD = 1.00) significantly higher than the Baseline condition (M = 5.05, SD = 1.64; W = 2.5, $p$ <.01) in terms of contributing different views and strengths to the ideation process. These quantitative findings indicate that participants more often framed the multi-agent system as a team of collaborators rather than a single tool.
Echoing the survey results, rather than relying on a single expert voice,  participants described the system as a collection of distributed roles that encouraged them to participate more actively in the interview. As P3 reflected, “it actually encourages me to join in the thinking process… like one of the people in the conversation,” while P19 valued how “[AI colleagues'] contributions complement each other.” This role separation often gave participants the impression of working with a team of specialists, with P10 noting that the role-playing “makes you feel more that everyone is contributing their strength.” 

At the same time, participants recognized trade-offs in this distributed setup. Some noted that AI colleagues tended to remain bounded within their professional domains, with P2 observing that “they’re limited to their professional aspects,” and P18 describing the system as expanding by “giving you another island” rather than filling out an existing territory. This metaphor reflected how colleagues generated separate contributions without consolidating expertise and broader topical coverage. By contrast, participants characterized the Baseline condition’s responses as “more academic and useful with wider scope” (P2) and emphasized its ability to fill in missing details to make ideas more complete (P18). In this sense, Baseline offered more integrated coverage within a single agent, while MultiColleagues emerged from the aggregate of multiple narrower perspectives, offering breadth across roles.

\subsubsection{Colleague Roles Enable Facilitative Leadership.} 
Findings from the interview showed that MultiColleagues generated facilitative leadership through distributed expertise. Since understanding multiple AI colleagues' voices required oversight and integration, participants often stepped into coordinator or facilitator roles. P3 explained that it “encourages me to join in the thinking process… like one of the people in the conversation,” while P5 reflected, “I really feel like I’m in front of this team, and they have to deliver their ideas to me, so I might feel that my sense of power is a little higher.” Importantly, this authority felt collaborative rather than authoritarian, with P16 noting, “I felt somewhere between leader, facilitator, but I definitely had the feeling of a collaborator.” The distribution of roles across AI colleagues demanded that users manage the collaborative process, as P17 described: “feel like chatting with my team in Slack, I can freely let whoever I want to speak out. I have a role feeling — like a PM [project manager].” The Baseline condition, in contrast, consolidated expertise into a single authoritative voice, reinforcing structured authority relations. Some participants described it as “a very senior-level expert who can immediately give you a very complete, very detailed plan” (P3), while others positioned themselves as supervisors with an assistant: “I feel like a boss. I’m just asking my assistant to fetch something for me” (P19). Yet even in supervisory roles, participants still noted how the Baseline condition shaped their thinking, with P12 reflecting, “I feel like I’m the boss… but [Baseline] reshapes how I think, so it has more power.” These accounts illustrate how MultiColleagues’ distributed roles fostered facilitative leadership, while Baseline’s consolidated expertise reinforced stable hierarchical structures.

\subsubsection{Team-Like Atmospheres Shape Collaborative Experience.}
Survey results indicated that the distribution of roles in MultiColleagues fostered a stronger sense of team-like interaction, with participants rating MultiColleagues (M = 5.75, SD = 1.02) significantly higher than the Baseline condition (M = 5.05, SD = 1.15; W = 17.5, $p$ = .046). These results suggest that the system’s role differentiation encouraged participants to experience the interaction as more socially collaborative.
Interview reflections further highlighted how role differentiation created a team-like dynamic. P5 noting that “everyone has their own role… it’s a very social state” and P10 adding that role-playing “makes you feel more that everyone is contributing their strength.” Others described a heightened sense of immersion, as P19 reflected that the team atmosphere “naturally facilitate[d] or control[led] the direction of the discussion.” For a few, this immersion was sufficiently strong to diminish the perceived boundary between human-AI contributions, with P9 recalling, “I felt like I forgot they were AI [colleagues].” 
By contrast, interview accounts of the Baseline condition consistently depicted it as neutral and instrumental. Participants described it as a source to “seek an answer and take the answer away” (P5). P9 echoed this perspective, “When I use it, I am a human being and [Baseline] is just a tool. I don’t feel it is my teammate.” Overall, while the Baseline condition was regarded as efficient and authoritative, MultiColleagues’ role differentiation cultivated stronger social immersion, reinforced the sense of presence, and encouraged more collaborative engagement.

To examine whether this heightened sense of team-like atmosphere was also reflected in participants’ own language, we analyzed participants' input through linguistic cohesion metrics from \textit{Coh-Metrix} \cite{graesser2004cohmetrix, mcnamara2014cohmetrix} alongside pragmatic style ratings across conditions (see detailed methods in Appendix \ref{appendix:b}.1). Results presented in Table~\ref{tab:metrics_comparison} showed a significant difference in Directness, with the Baseline condition (M = 5.82, SD = 0.91) rated significantly higher than the MultiColleagues condition (M = 5.21, SD = 0.88; W = 30.0, $p$ = .009), suggesting that interactions with Baseline elicited more straightforward, task-focused language. Other contrasts did not reach statistical significance.

\begin{table*}[ht]
\centering
\caption{Statistical Comparison for Performance \& Integration Metrics between \textbf{MultiColleagues (MC)} and \textbf{Baseline}. Results show significantly higher self-reported ratings for MC on \textit{Creative Exploration} and \textit{Perceived Quality \& Novelty}.}
\begin{tabular}{lcccccc}
\toprule
\textbf{Metric} & \textbf{Q\#} & \textbf{MC (M ± SD)} & \textbf{Baseline (M ± SD)} & \textbf{W} & \textbf{p-value} & \textbf{Effect Size ($r$)} \\
\midrule
Creative Exploration & Q1, Q4 & 6.00 ± 0.87 & 4.95 ± 1.55 & 31.5 & \textbf{.018*} & 0.63 \\
Process Enrichment & Q3 & 5.80 ± 1.20 & 5.00 ± 1.72 & 22.5 & .054 & 0.45 \\
Perceived Quality \& Novelty & Q5, Q7 & 5.95 ± 0.92 & 4.97 ± 1.16 & 17.0 & \textbf{<.01**} & 0.69 \\
\bottomrule
\end{tabular}
\label{tab:rq1_revised}
\end{table*}

\subsubsection{Engagement and Flow Enhance Collaborative Immersion}
Participants reported experiencing stronger engagement and conversational flow in MultiColleagues, with significantly higher ratings in the MultiColleagues condition (M = 5.70, SD = 1.38) compared to the Baseline condition (M = 4.45, SD = 1.57; W = 30.0, $p$ = .014). Behavioral interaction measures further reinforced this pattern (Table \ref{tab:user_interaction}). Participants contributed nearly twice as many utterances in MultiColleagues (M = 8.35, SD = 5.79) than in the Baseline condition (M = 4.10, SD = 2.45; W = 9.5, $p$ = .001) and produced substantially more words overall (W = 18, $p$ < .001), suggesting an increased engagement in ideation in the MultiColleague condition compared to the Baseline. Sessions also lasted longer in MultiColleagues (M = 12.90, SD = 2.80) compared to Baseline (M = 9.80, SD = 2.30; W = 33, $p$ = .006), suggesting that participants remained more cognitively and temporally invested when coordinating multiple voices. 
Interview reflections echoed these dynamics. P5 explained, “I feel like in my discussions, I’m constantly prompting new information and asking for new information, kind of wanting to lead the discussion. I feel I’m more like a facilitator.” Others emphasized how the incremental delivery kept them attentive and invested (P3, P10, P12). While MultiColleagues fostered ongoing involvement, Baseline condition’s one-shot responses were often described as efficient but less engaging. P19 described the interaction as “relatively one-way, like a presentation… I just pick what I want,” and P8 admitting, “I didn’t feel like we were collaborating… I just wanted to hear its perspective.” These findings show that MultiColleagues’ structured rhythm sustained deeper immersion and engagement, while Baseline delivered efficiency at the cost of shallower participation.

\subsection{RQ2: Perceived Impact on Ideation Process and Outcome Quality}
To address RQ2, we examined participants’ evaluations of creative outcomes along three dimensions: creative exploration, process enrichment, and perceived outcome quality and novelty. Participants rated MultiColleagues more favorably than Baseline across all three dimensions, with significant differences for creative exploration and perceived outcome quality and novelty (Table~\ref{tab:rq1_revised}).

\begin{figure*}[t]
    \centering
    \includegraphics[width=0.9\linewidth]{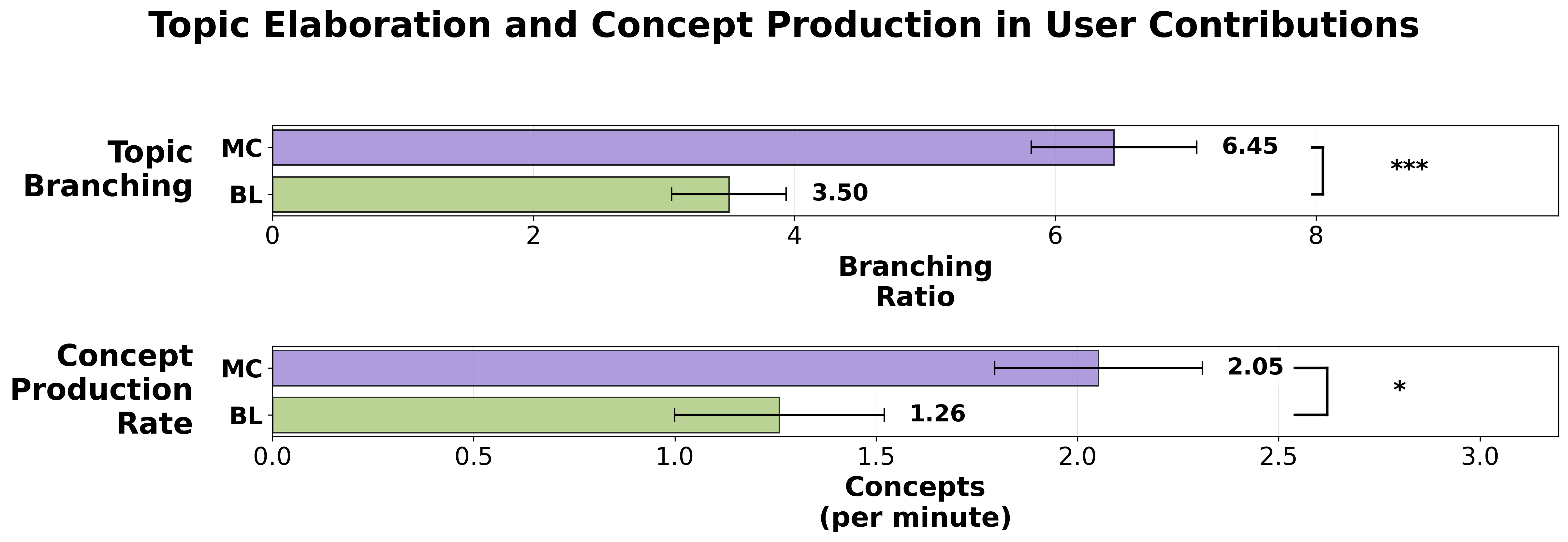}
    \caption{This figure compares users' idea development patterns across MultiColleagues (MC) and Baseline (BL). MC produced a higher topic branching (6.45 vs. 3.50, $p < 0.001$) and higher concepts production rate (2.05 vs. 1.26 concepts per minute, $p < 0.05)$.}
    \Description{Statistical comparison of users' ideation development between Baseline and MultiColleagues. The figure shows that MultiColleagues demonstrate significantly more topic branching and concept production, which supports broader and more in-depth ideation.}
    \label{fig:conversational_topics}
\end{figure*}

\subsubsection{MultiColleagues Broadens Creative Exploration.}

Survey measures confirmed that participants experienced greater creative exploration with MultiColleagues (M = 6.00, SD = 0.87) than with the Baseline condition (M = 4.95, SD = 1.55; W = 31.5, $p$ = .018). Further, statistical tests did not reveal significant associations between number of personas selected and creativity scores ($r=0.223, p=0.345$) or occupational background ($H=1.292, p=0.524$). These findings suggest that inclusion of a good enough number of different perspectives would facilitate creativity; however, increasing the number of personas solely did not guarantee higher creativity. To further examine this difference, we analyzed the topical structure and conceptual content of user contributions. Topic branching captured how many parallel idea directions participants pursued, while concepts production reflected the amount of concepts (i.e., nouns, \cite{kintsch1998comprehension}.) people produced in their language discourse.
Figure~\ref{fig:conversational_topics} shows that MultiColleagues supported broader and more in-depth ideation than the baseline. Participants generated significantly higher topic branching with MultiColleagues (M = 6.45, SD = 2.54) than Baseline (M = 3.50, SD = 1.79, $t(19) = 5.15, p<.001$), indicating that they explored more parallel lines of ideas in terms of broader elaboration of the topics. Participants also generated more concepts in their discussion, suggesting more in-depth reflections, as shown in a higher concept production rate (M = 2.05 concepts/min, SD = 0.81) in MultiColleagues than in Baseline (M = 1.26, SD = 0.67, $t(19) = 2.59, p<.05$). These results show MultiColleagues encouraged both greater structural divergence and richer conceptual expression during ideation.
Interview reflections reinforced this pattern. Several participants noted that the Baseline condition’s “large initial response” often felt like “a presentation you just read” (P19), which reduced their motivation to extend the ideas in breadth or depth. In contrast, MultiColleagues encouraged participants to “proactively join to think and discuss” when only a few points were offered (P3). These results demonstrate that MultiColleagues broadened creative exploration by enabling participants to pursue more directions and engage in more conceptually rich discourse.

\subsubsection{Traceable Perspectives Enable Structured Integration.}
We found from interviews that the traceability of perspectives in MultiColleagues supported participants’ ability to integrate ideas. Because contributions were anchored to distinct AI colleagues, participants reported it was easier to follow up, remember, and combine ideas into coherent outcomes. P4 reflected that hearing “different angles, different perspectives” made the information “stay in your mind rather than just flashing by,” while P3 emphasized  the system “divides into several colleagues… when I want to go deeper, I clearly know which one to talk to.” These role-based distinctions provided a clear map of where ideas originated, helping participants balance and integrate perspectives into more structured outputs. In contrast, the Baseline condition merged perspectives into a single response. While this blending reduced the ability to trace individual contributions, it benefited comprehensive coverage, offering “extensive lists and complete answers” when participants wanted a consolidated view (P2, P4, P6, P19).

\subsubsection{Conversational Rhythm Shapes Perceived Idea Quality and Novelty}
Self-reported measures indicated that participants perceived their ideas as higher-quality and more novel with MultiColleagues, with ratings significantly higher in the MultiColleagues condition (\textit{M} = 5.95, \textit{SD} = 0.92) than in the Baseline condition (\textit{M} = 4.97, \textit{SD} = 1.16; \textit{W} = 17.0, $p$ < .01).

To complement these perception-based measures, we additionally examined idea originality using a TTCT-inspired scoring approach \cite{guzik2023originality, chi2024evolutionary, hadas2024using, fukumura2025creativity}. 
TTCT evaluates creativity across fluency, flexibility, originality, and elaboration. In this study, fluency and flexibility were not included, as previous works demonstrated that LLMs tend to produce a high volume of responses that artificially inflate fluency scores, while flexibility is strongly confounded by fluency and thus offers limited validity as an independent measure \cite{guzik2023originality,chi2024evolutionary,hadas2024using}. Instead, we focused on originality as the core dimension. Following prior work \cite{fukumura2025creativity}, originality ratings were generated using GPT-5 with the same validated evaluation prompt applied to participants’ full conversations. Each conversation was evaluated on a standardized 5-point rubric, with three independent runs per idea set, and the averaged score was used for originality analysis. The results showed a trend that MultiColleagues (\textit{M} = 3.78, \textit{SD} = 0.38) scored higher than Baseline (\textit{M} = 3.59, \textit{SD} = 0.44; \textit{W} = 70.0, $p$ = .202), but this difference was not statistically significant, diverging from the perceived novelty reported by participants. Interview accounts help explain this divergence by highlighting differences in the rhythm of idea generation. Participants emphasized that MultiColleagues generated ideas progressively, in a rhythm they described as “digestible… aligned with the rhythm of human discussion” (P19). This stepwise unfolding supported a process of guided discovery, where ideas “emerge through guided conversation” (P2) and developed like a “chain of thought” (P16). In contrast, Baseline often produced an immediate burst of ideas. Participants acknowledged these outputs as “very creative right from the start” (P2), yet also noted that density could feel overwhelming. Several reported this rapid surge made it harder to refine and act on ideas, compared with MultiColleagues’ more deliberate, step-by-step approach (P3, P5, P15).

\subsubsection{Breadth–Depth Trade-offs Shape Outcome Enrichment.} 
Survey measures suggested that outcome enrichment was marginally stronger in the MultiColleagues condition (\textit{M} = 5.80, \textit{SD} = 1.20) compared to the Baseline condition (\textit{M} = 5.00, \textit{SD} = 1.72; \textit{W} = 22.5, $p$ = .054). This pattern reflects a breadth–depth trade-off between two systems. 
Interview reflections described MultiColleagues as a breadth-first approach, expanding the solution space through multiple diverse perspectives. Participants valued its ability to enrich the early, exploratory stages of ideation. P1 noted that it was effective for “expanding ideas,” while P7 described its output felt “comprehensive” because it integrated “multiple angles [such as] logic and marketing,” thereby fostering cross-functional thinking. This breadth encouraged novelty and variety but often came at the cost of providing concrete, actionable details.
In contrast, the Baseline condition was viewed as a depth-first tool, excelling at producing more focused, polished, and immediately usable outcomes. P4 explained that it went “a step further” than MultiColleagues by providing tangible examples like “sample data,” which offered a “more concrete understanding of how to process a dataset.” Other highlighted its utility for “executive decision-making” (P6) and for “executing idea and goal” (P12). This depth and implementability gave Baseline an advantage in later-stage tasks requiring clarity and execution, whereas MultiColleagues dominated in the initial phase by maximizing creative possibilities. 


\subsection{RQ3: System Design Features and User Agency in Creative Ideation}
We evaluated how system design features influenced user agency (RQ3) during creative ideation across four metrics: user guidance, user control, adaptive thinking mode, and future use intent. As shown in Table~\ref{tab:rq3_agency}, MultiColleagues received significantly higher ratings than Baseline for user control and adaptive thinking mode.
\begin{table*}[ht]
\centering
\caption{Statistical Comparison for Agency \& Control Metrics between \textbf{MultiColleagues (MC)} and \textbf{Baseline}. Results show significantly higher ratings for MC on \textit{User Control} and \textit{Adaptive Thinking Mode}.}
\begin{tabular}{lcccccc}
\toprule
\textbf{Metric} & \textbf{Q\#} & \textbf{MC (M $\pm$ SD)} & \textbf{Baseline (M $\pm$ SD)} & \textbf{W} & \textbf{p-value} & \textbf{Effect Size ($r$)} \\
\midrule
User Guidance & Q2 & 5.85 ± 1.09 & 5.40 ± 1.39 & 34.5 & .436 & 0.23 \\
User Control & Q12 & 5.80 ± 1.32 & 4.40 ± 1.79 & 27.0 & \textbf{.033*} & 0.55 \\
Adaptive Thinking Mode & Q9 & 5.90 ± 1.29 & 4.60 ± 1.70 & 16.5 & \textbf{.023*} & 0.58 \\
Future Use Intent & Q11 & 6.15 ± 1.09 & 5.50 ± 1.61 & 11.5 & .098 & 0.38 \\
\bottomrule
\end{tabular}
\label{tab:rq3_agency}
\end{table*}

\subsubsection{Autonomous Versus Manual Direction Shape User Guidance.} 
Participants’ ratings on user guidance indicated no significant difference between MultiColleagues (\textit{M} = 5.85, \textit{SD} = 1.09) and Baseline conditions (\textit{M} = 5.40, \textit{SD} = 1.39; \textit{W} = 34.5, $p$ = .436). 
Nonetheless, interview findings revealed that the two systems supported user guidance in distinct ways. MultiColleagues was described as more autonomous, with participants noting that once a question was posed, the system could sustain its own line of discussion. Participants explained, “I can throw out a question, then they start an intense discussion” (P13, P15). This process was perceived as unfolding like a “chain of thought” (P16), where initial contributions created openings for participants to engage selectively. As P3 explained, “It gives you 1–2 points first, so you will proactively join to think and discuss”. However, this autonomy also carried drawbacks, as conversations sometimes drifted from the intended focus and required effort to redirect (P12, P13). 
In contrast, interview accounts of the Baseline condition highlighted its reliance on manual direction. Participants described it as “question-answer, question-answer… I have to tell it what I want to do next for each step” (P13). While this approach offered precise control, it was effort-intensive and likened to “guid[ing] it in a very formal way, just like prompt engineering” (P16). Its polished and comprehensive outputs could also constrain participation, leaving some participants “lost, don’t know how to chat or continue” (P3).

\subsubsection{MultiColleagues Empowers Participants with Greater Control} 
Participants reported a significantly stronger sense of control when working with MultiColleagues (\textit{M} = 5.80, \textit{SD} = 1.32) compared to Baseline (\textit{M} = 4.40, \textit{SD} = 1.79; \textit{W} = 27.0, $p = .033$). 
Interview reflections provided further insight into this perceived control through two main factors. First, participants attributed the enhanced control to MultiColleagues’ support for shifting between explore and focus thinking modes. Participants could “move from exploring new ideas to working on a specific one” (P9) and “control it if I don’t want it to be divergent” (P19). In contrast, the Baseline condition offered no such flexibility, leaving participants feeling they were “always fixing its direction” (P9). However, some valued its chunked, turn-by-turn dialogue, which created opportunities to “chime in to change direction at any time (P19).” 
Second, participants associated greater control with the system's alignment to their own goals. MultiColleagues was described as “more engaged, better aligned with my original direction” (P12), and even supported leadership skills, as P16 noted, “I was able to bring in other agents when they were being quiet, which is actually a great team leadership learning experience.” By contrast, Baseline was seen as “strong, professional, very dominant” (P3) to overshadow participants' intent. 

At the same time, participants in interviews also acknowledged trade-offs in managing multiple colleagues' voices. Unlike Baseline, which was described as linear and predictable (P6), MultiColleagues required greater coordination effort. Participants pointed out that the diversity of roles occasionally caused drift, requiring “firm willpower to keep this conversation stable” (P6) or effort to pull AI colleagues “back on track” (P12, P13). P8 also described subtle “social pressure” when navigating overlapping perspectives. Overall, we found that MultiColleagues offered flexible and participatory control but demanded coordination, whereas Baseline provided predictable yet more rigid control.

\subsubsection{Adaptive Thinking Mode Strengthens Flexibility in Ideation}
Survey results showed that participants rated MultiColleagues (\textit{M} = 5.90, \textit{SD} = 1.29) significantly higher on adaptive thinking mode compared to the Baseline condition (\textit{M} = 4.60, \textit{SD} = 1.70; \textit{W} = 16.5, $p = .023$), indicating stronger support for shifting between exploration and focus during ideation. 
Interview reflections further explained this flexibility. Participants emphasized MultiColleagues’ ability to deliberately transition between divergent and convergent thinking, which they saw as central to managing the creative process. They described being able to “move from exploring new ideas to working on a specific one” (P9), and “control it if I don’t want it to be divergent” (P19). Facilitation further supported these shifts by prompting reflection and offering lightweight guidance without imposing direction. Participants valued being asked whether to “dive deeper” or “explore other” directions (P15, P7), which helped them regulate attention and decide when to transition. As P6 described, the facilitator acted more as “a guide,” providing reminders and summaries that supported concentration without steering outcomes. 
Participants also highlighted the value of explicit, manual controls for shifting modes. Button-based toggles made transitions quicker and more natural within the flow of ideation, reducing the need to type additional instructions. As P19 described, “the quickest way is to click to switch,” while P20 noted that visible controls were preferable because “I don’t have to type so many words, I can just click.” Beyond convenience, this design also provided a structural trace of shifts, helping participants track how their workflow moved between breadth and depth.

\begin{figure*}[h]
    \centering
    \includegraphics[width=1.0\linewidth]{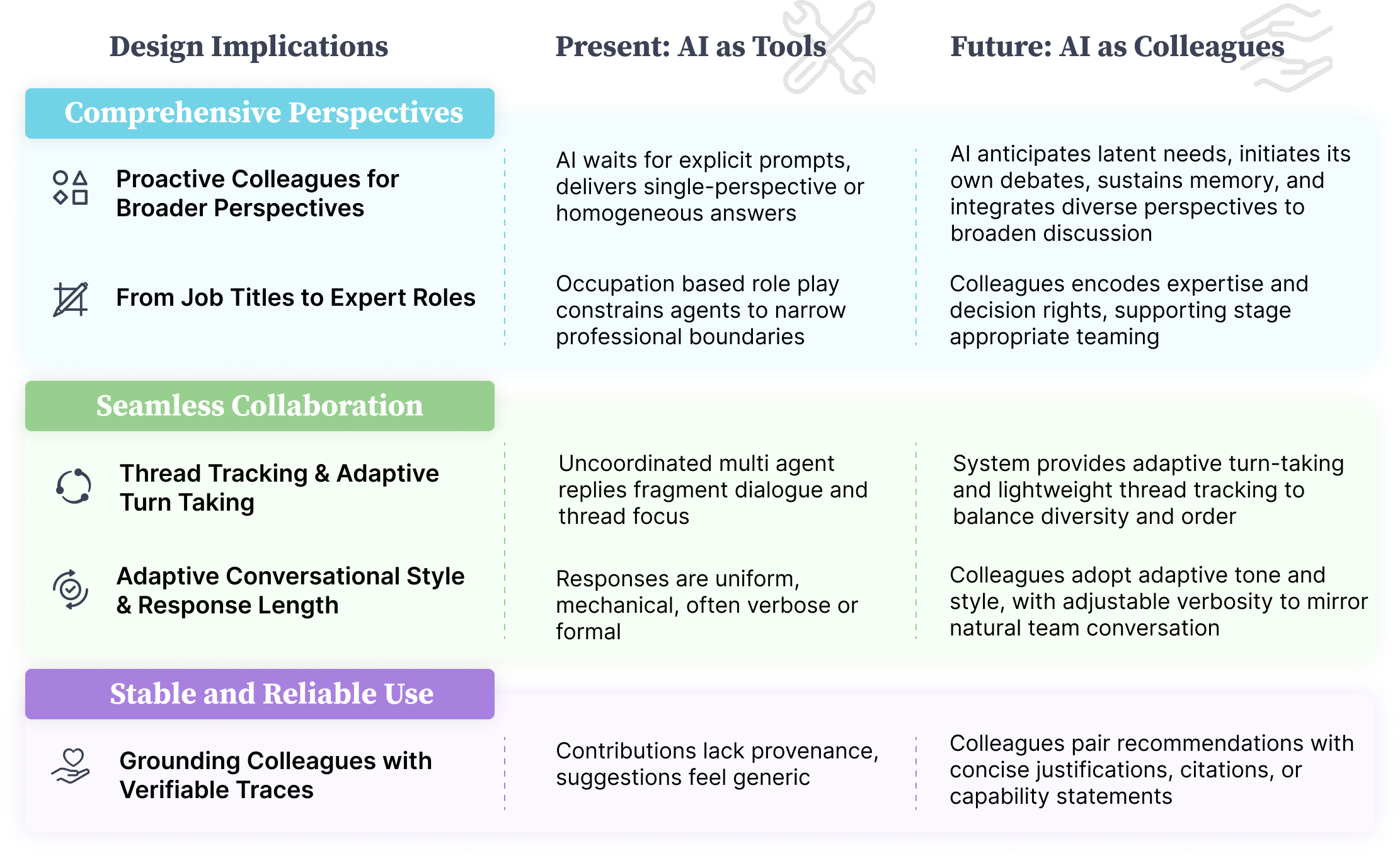}
    \caption{\textbf{Design Implications for Human–Multi-Agent Collaboration: From AI as Tools to AI as Colleagues.} This table summarizes six design implications that guide the transition from single-agent, tool-like AI to multi-agent colleagues in collaborative ideation. Each implication highlights a key design consideration (left column) and contrasts current AI as task-oriented tools (middle) with envisioned dynamics of human–multi-agent collaboration, where AI act as proactive, accountable, and socially embedded colleagues (right).}
    \Description{Design implications for human–multi-agent collaboration. The figure presents 5 implications that mark the shift from AI as task-oriented tools to AI as proactive, accountable, and socially embedded colleagues, highlighting considerations in agents' proactivity, persona dimensions, orchestrations, conversational style, and reliability.}
    \label{fig:DI}
\end{figure*}

\subsubsection{Future Use Intent Depends on Task Stage and Context.}
Survey results reflected participants' high intent to use both systems, with MultiColleagues (\textit{M} = 6.15, \textit{SD} = 1.09) rated slightly higher than the Baseline condition (\textit{M} = 5.50, \textit{SD} = 1.61; \textit{W} = 11.5, $p$ = .098), though the difference was not statistically significant. 
Interview reflections further revealed how adoption was shaped by task stages and context. Many participants described a staged workflow in which MultiColleagues was used early to generate and expand ideas, followed by the Baseline condition to validate, refine, or translate those ideas into actionable steps (P6, P10). For example, P6 explained they would “start from [MultiColleagues] to find good new ideas, then bring this idea to [Baseline]” for detailed implementation, while P10 planned to “discuss a few ideas first” with MultiColleagues and then use Baseline to analyze feasibility and trade-offs. 
Besides, the suitability of each system was also tied to problem clarity. MultiColleagues was seen as particularly useful when questions required compound perspectives or when participants sought to brainstorm from multiple angles (P8, P18, P20). By contrast, the Baseline condition was considered more efficient and specific for narrow, practical, or well-defined tasks (P13, P15, P17). 
Participants also noted limitations during interviews. MultiColleagues was considered more suitable for large or complex problems that benefit from multiple perspectives (P7), ut some highlighted challenges such as a steeper learning curve and the need for longer engagement to fully realize its value (P9). These accounts suggest that future use is less about preferring one system overall and more about strategically aligning each with the stage, scope, and complexity of the problem at hand.

\section{Discussion} 
\label{dicussion}

\subsection{Summary of Results}
Our study demonstrates how MultiColleagues, a multi-agent conversational system, reshapes collaborative ideation processes compared to a single-agent baseline. First, for \textbf{RQ1 - Collaborative Experience}, MultiColleagues fostered a distributed, team-like atmosphere. Participants reported stronger team-like feelings and complementary strengths across roles. Engagement also increased, with nearly twice as many utterances, longer sessions, and more words produced overall, reflecting a shift from passive receipt to facilitative coordination. For \textbf{RQ2 - Perceived Creative Outcomes}, the exposure to multiple AI colleague perspectives supported broader and deeper exploration, participants spent more time developing each idea topic and reported higher perceived outcome quality and novelty. For \textbf{RQ3 - System Design and User Agency}, MultiColleagues enabled stronger perceived control and more adaptive thinking mode support for switching between divergent and convergent thinking. Although autonomy sometimes led to conversational drift, participants valued the ability to steer discussions, regulate rhythm, and align outputs with evolving goals. Our results indicate that multi-agent systems shift ideation from tool use toward dynamics that resemble collegial teamwork. Building on these findings, we outline key design implications that translate observed collaboration patterns into actionable directions for future system design, detailed in the following sections (Figure \ref{fig:DI}).

\subsection{Expanding Comprehensive Perspectives through Proactive Multi-Agent Collaboration}

Our study demonstrated how MultiColleagues broadened user perspectives by combining multiple voices with proactive contributions. Instead of passively awaiting prompts, colleagues anticipated latent needs and introduced dimensions users had not explicitly considered, thereby helping them move beyond current frames of thought. 
Building on this, we frame MultiColleagues system as a foundation for multi-voiced conversation, where diverse AI colleagues contribute proactively and in contrast to one another, creating exchanges that are both cognitively supportive and closer to real teamwork.

\subsubsection{Design Implication 1: Proactive Colleagues Open New Perspectives}

Prior work has noted the risks of homogeneous perspectives in AI systems, while also emphasizing the potential of multi-agent conversation to counteract them by surfacing diverse viewpoints and stimulating creative exchanges \cite{zhang2024filterbubbles, pang2025synthetic, lu2024llm}. Our findings extend this work by showing that users valued not only the presence of multiple agents but also their proactive behavior. Colleagues that raised overlooked considerations or challenged premature convergence helped broaden perspectives and sustain engagement \cite{zhang2021ideal}. 
In our prototype, each colleague followed role based instructions to monitor the discussion and intervene when they could add a new angle or critique. Colleagues also contrasted and compared one another’s ideas, which allowed alternative viewpoints to surface and be considered before reaching the user. We also designed the  facilitator who served as the main conversational lead, tracking how the conversation unfolded and indicating when the group should continue exploring or shift toward refinement.

Future systems could introduce more proactive colleagues that recognize when and how to take initiative in the discussion. Colleagues and facilitators could label the function of their contributions, for instance by marking a message as opening a new issue or responding with a challenge to a prior idea. By shaping when colleagues and facilitators intervene, and how their perspectives are contrasted and summarized, designers can better harness proactive multi-agent collaboration as a resource for broadening perspectives rather than overwhelming users.

\subsubsection{Design Implication 2: From Occupational Role-play to Domain Expert Teamwork}

In this study, personas were primarily differentiated by occupation, which provided distinct perspectives but offered limited guidance on how colleagues should contribute across exploration and evaluation stages. Interviews indicated that this occupation focused framing sometimes kept colleagues within narrow professional boundaries and left participants uncertain about who had sufficient expertise or authority for specific issues (P2, P7, P18). 
P7 explicitly noted that occupational breadth was not always the most relevant form of diversity. When his discussion focused on a specific machine learning subproblem, bringing in more cross occupational colleagues did not help his ideation. Instead, he emphasized the value of differentiating colleagues within the same domain by specialization focus and level of technical authority so that some colleagues could provide rigorous critique and others could take responsibility for synthesis during convergence.

Recent multi-agent ideation work also finds that increasing the number of agents is insufficient. Idea quality depends on agents with adequate expertise who can sustain critique and refinement over time \cite{beyondbrainstorming2025}. For system design, this implies encoding clearer differences in seniority and decision rights among colleagues, so evaluation feedback is appropriately confident and responsibility for synthesis is unambiguous. Work on conversational personas and framing shows that persona cues and AI's presented role can shape user expectations and interaction strategies, therefore should be designed to avoid deceptive signaling or reinforcing stereotypes \cite{googlepersona2018, liao2021roleframing}. Future systems could treat colleague configurations as explicit, adjustable dimensions and examine how they shape teammate selection, for example preferring a smaller set of high authority evaluators for convergence and adding more contrasting colleagues for early stage exploration \cite{beyondbrainstorming2025}.

\subsection{Orchestrating Many Voices: Designing Multi-Agent Colleagues for Seamless Collaboration}
While multi-agent autonomy offers opportunities for richer perspectives, it also raises challenges of coordination and control. Our MultiColleagues system demonstrated how multi-agent conversation supports idea integration by making perspectives traceable and diversifying problem-solving. Unlike single-model interactions that remain sequential and Q\&A-driven, MultiColleagues exchanges naturally introduced complementary viewpoints, encouraging participants to synthesize across roles. The progressive orchestration of colleagues promoted smoother collaboration by breaking down problems into multiple angles and guiding participants toward more deliberate integration of ideas.

\subsubsection{Design Implication 3: Initiating Thread Tracking and Adaptive Turn Taking in Multi-Agent Chat.}

Managing contributions from multiple AI colleagues requires orchestration strategies that surface diverse perspectives without overwhelming users. Prior work highlights approaches such as round-table settings, sequential workflows, phased role-play, and chat-based interface for balancing diversity and order in multi-agent systems \cite{pang2025synthetic, lu2024llm, dibia2024autogen, peng2024navigatingunknownchatbasedcollaborative}. Our study extends this line of work by showing that a conversational chat style with clear turn taking lowered barriers to entry and preserved a distributed sense of expertise, which in turn produced a digestible rhythm that supported engagement and coordination during short, time bounded ideation sessions.

As the number of AI colleagues grows, legibility and fast attribution become central. Users need to “know which one to talk to” (P3), follow the development of each thread, and return to specific contributions when ideas converge. In our prototype, distinct avatars and role labels prompted participants to treat colleagues as recognizable teammates rather than a single undifferentiated system voice, which aligns with research that links visual embodiment to stronger social presence and coordination in group work \cite{shamekhi2018facevalue, hilpert2024avatarvisualsimilaritysocial}. Building on this, multi-agent interfaces could pair identity cues with lightweight visualization, such as color coded dialogue flows or labeled discussion maps, thus users can trace how proposals and critiques evolve.
Participants also imagined more flexible and controllable interaction patterns. They described needs such as reconfiguring colleagues mid discussion (P1, P5), inviting simultaneous responses for comparison (P2), and setting up focused debates between evaluators and proponents (P4, P7). These expectations point toward identity based steering hooks where users can click a colleague to request targeted critique, re-invite a silent expert, mute a dominant voice, or change who speaks first when the team moves from exploration to convergence.

From the system side, our prototype used prompt level instructions to coordinate turns. Each colleague received the shared conversational history and was asked to decide whether to speak in the current round. The orchestrator then ranks the colleagues with higher reported priority for each turn that produces a manageable flow.
Beyond this basic scheme, adaptive orchestration can draw on a richer set of conversational signals. Prior works show that interaction traces and user cues can guide when agents take or yield the floor \cite{hu2025dialoglab, xu2013engagement}. Building on this, multi-agent systems can monitor which roles participate and how often users seek clarification, then adapt the mix of colleagues and the pacing of turns to fit the current phase of work. Future work could formalize such adjustments as phase sensitive orchestration policies, extending existing models of turn taking and turn detection in multi-party systems. \cite{Aldahoul2024AIgeneratedFI}.

\subsubsection{Design Implication 4: Adaptive Conversational Style and Response Length Facilitate Socialized Interactions}
The communicative style of AI colleagues strongly influenced how smoothly information was exchanged and understood. Users' interactions felt closer to peer-to-peer collaboration rather than mechanical delivery when AI responses adopted a more approachable and adaptable tone that could be polite, occasionally humorous, or appropriately formal. Subtle stylistic shifts created a sense of social presence across conversations that resemble real workplace exchanges and making shared content easier to follow. In addition, response length also shaped the rhythm of collaboration. Participants reflected that longer outputs were described as valuable in “focus mode” to offer more in-depth reasoning, while shorter and contextualized statements helped anchor contributions without overwhelming the discussion (P20). Future works could focus on designing adaptive mechanisms that vary in verbosity across phases of a task, from elaboration at the outset to concise handovers later, that can support both breadth–depth exploration and efficient knowledge sharing.

\subsection{From “Many Agents” to Colleagues: Role Stability and Reliable Use}

Building on multi perspective proactivity (Section 6.2) and seamless orchestration (Section 6.3), a further step is needed before AI can be treated as colleagues rather than tools. Users need to see agent contributions as coming from identifiable partners whose ideas are stable enough and reliable enough to reuse. Our study suggests that interacting with multiple differentiated agents can shift this perception. Compared with the single agent baseline, MultiColleagues fostered a stronger sense of “being in a meeting” with a team. Participants described the experience as “hosting a meeting” where different colleagues contributed ideas and took on recognizable roles. Some noted that the system felt “between a tool and a teammate” (P9), capturing a transitional stage from instrumentality to collegiality. The stability of each colleague’s behaviour and the diversity of perspectives encouraged participants to revisit AI suggestions and weave them into their own ideas, rather than treating outputs as one off prompts (P20). This pattern is consistent with recent work on multi-agent conversational systems, which finds that structured role distributions and visible participation can support social presence and team like collaboration \cite{sun2025personal, pang2025synthetic}. At the same time, participants also questioned when colleagues were actually knowledgeable, which points to the need to complement social framing with better grounding of what personas say.

\subsubsection{Design Implication 5: Grounding Multi-Agent with Verifiable Traces for Reliable Use} Our findings suggest that participants are more willing to treat persona contributions as worth considering when they can see why a suggestion was made and what it is based on. Role playing alone is insufficient. When responses lacked visible grounding, some participants experienced them as generic or performative rather than as input from a knowledgeable peer. This aligns with work showing that people engage more carefully with AI advice when they can inspect reasoning traces or provenance and decide when to accept or discount a suggestion \cite{yang2023harnessingtrust, hoque2024hallmark, lee2025llm}. 

To support reliable use, multi-agent systems should couple colleague style interaction with verifiable traces, such as inline references, source tags, or concise capability statements \cite{yang2023harnessingtrust}. Such cues help users judge reliability on fluent but unfounded content. Future work can explore pipelines that combine persona prompts with retrieval augmented generation or domain tuned models so that AI colleagues are not only experienced as a team, but are also grounded enough to be integrated into users’ reasoning in practice.

\subsection{\textbf{Ethical Considerations}}
The move toward AI as colleagues also raises significant ethical questions. As AI systems increasingly contribute to creative work, scientific discovery, and decision-making, the boundaries of authorship and accountability become less clear. For example, recent debates on AI credited as co-authors highlight both the promise and tension of expanding machine agency in knowledge production \cite{hugenholtz2021copyright, ateriya2025exploring}. Beyond authorship, using AI generated personas introduces additional considerations. Empirical studies show that LLM personas often appear coherent and convincing \cite{schuller2024llmpersonas,smrke2025exploring}, yet they also encode systematic demographic and cultural biases \cite{salminen2024deus} and can subtly reinforce stereotypes if deployed uncritically. Even when personas are derived from real datasets, model-level variation and identity drift necessitate mechanisms for verification and correction \cite{jung2025personacraft}.
These risks underscore the importance of transparency, controllability, and user agency in multi-agent systems. As AI colleagues autonomously advance discussions and generate new perspectives, mechanisms are needed to ensure that their contributions remain interpretable and that humans retain meaningful influence over outcomes. Without such measures, agentic AI risks creating “moral crumple zones” where responsibility becomes diffused and no actor is fully accountable \cite{mukherjee2025agentic}. Designing for auditability, feedback integration, and human oversight will therefore be critical. The collegial paradigm cannot be achieved without parallel efforts to establish ethical frameworks that safeguard accountability while enabling AI to act as trusted partners in collaborative processes.

\subsection{Limitations and Future Work}
While our study offers initial insights into AI colleagues, several limitations constrain the scope of our claims and suggest directions for future work. 
First, our participant pool consisted mainly of students and early career professionals. Expectations of teamwork and authority can vary across domains, seniority levels and cultural contexts, so future work should recruit more diverse samples to examine how people in different professions and regions integrate AI colleagues in everyday collaboration. 
Second, participants' interaction was short and limited to a single ten minute session. Longer term deployments are needed to understand how AI colleagues support full cycles of exploration, debate and convergence.

The system design also introduces important constraints. All personas were instantiated from the same language model, which limited diversity of perspectives and sometimes produced similar response styles. Future work should investigate heterogeneous model setups and mechanisms that enforce divergent stances or knowledge sources. Besides, we studied a relatively small and short lived team of nine colleagues, whereas real workplaces often involve larger and more persistent teams. Subsequent systems should examine orchestration strategies that keep multi agent collaboration coherent as the number of colleagues and the duration of interaction increase.

A further limitation concerns our use of an LLM-based judge to evaluate idea originality. While this approach, following prior work, enables scalable and consistent scoring, independent validation on the alignment between LLM-generated ratings and human judgments has not been performed. As LLM-based evaluation remains an emerging practice, future work should triangulate originality assessments with human raters or mixed evaluation protocols to better establish validity.

Finally, persona fluency and inherited biases pose risks for reliable use. Participants at times treated confident answers as knowledgeable even for underspecified or difficult questions, echoing findings that LLM generated ideas may appear promising yet degrade under closer scrutiny \cite{si2024llmsgeneratenovelresearch, si2025ideationexecutiongapexecutionoutcomes}. Personas also inherit demographic and cultural biases from underlying models \cite{salminen2024deus, smrke2025exploring}, and in multi agent settings they may reinforce one another’s assumptions or respond confidently to unanswerable questions \cite{kaate2025you}. Future systems should integrate bias auditing and uncertainty signaling, and provide mechanisms that help users critically evaluate persona contributions, especially in high stakes or sensitive domains.

\section{Conclusion}
To investigate how AI might move beyond functioning as tools toward acting as peer-like colleagues in collaborative contexts, we developed MultiColleagues, a system that orchestrates multiple role-differentiated personas and incorporates facilitation and thinking-mode features to support structured ideation development. In a within-subjects study that compared MultiColleagues with a single-agent baseline, we found that MultiColleagues was associated with higher perceived outcome quality and novelty, a stronger sense of team presence, and more active engagement with greater interaction control.  Building on these findings, we derived several design implications for amplifying proactive contributions, enabling more seamless human–agent coordination, and supporting calibrated trust. Taken together, this work advances the growing body of research on multi-agent systems by showing how design choices shape whether users perceive AI as tools or as peer-like collaborators. We hope these insights will inform future research and design efforts aimed at creating generative multi-agent systems that not only enhance human creativity but also lay the groundwork for trustworthy and sustainable forms of AI colleagueship.

\begin{acks}
We thank Professor ChengXiang Zhai and the TIMAN group for providing computational resources and valuable feedbacks that supported this work. We also thank all study participants for their time and contributions.
\end{acks}



\bibliographystyle{ACM-Reference-Format}
\bibliography{reference}

\appendix

\section{Appendix A}
\label{appendix:a}

\subsection {Pre-Survey Creativity Questionnaire}

Participants rated their agreement with the following statements on a 7-point Likert scale  
(1 = Strongly Disagree, 7 = Strongly Agree). The scale demonstrated excellent internal consistency (Cronbach’s $ \alpha = 0.956 $).

\begin{enumerate}
    \item I often come up with new and practical ideas to improve performance.
    \item I search for new technologies, techniques, or solutions.
    \item I suggest new ways to increase the quality of work or outcomes.
    \item I am a good source of creative ideas.
    \item I come up with creative solutions to problems.
    \item I often have a fresh approach to challenges.
    \item I am willing to take risks in generating new ideas.
    \item I promote and support ideas that I believe in.
    \item I create detailed plans for implementing new ideas.
    \item I exhibit creativity when given the opportunity.
    \item I consider myself a creative person.
\end{enumerate}

\subsection{ Post-System Evaluation Questionnaire}
Participants rated their agreement with the following statements on a 7-point Likert scale  
(1 = Strongly Disagree, 7 = Strongly Agree).

Q1. The system encouraged creative thinking and helped me explore a wide range of ideas.

Q2. I was able to guide the idea generation based on my needs during the task.

Q3. This system benefits/enriches my ideation process and thinking.

Q4. I reached lots of valuable or actionable ideas that felt better than what I might have generated alone.

Q5. Working with the AI system allows me to develop more creative solutions that I would not have come up with on my own.

Q6. Interacting with the system felt like working with a helpful teammate.

Q7. The system offered useful perspectives that expanded or deepened my thinking.

Q8. When working with this AI system, everyone (human/AI) can contribute their strengths and complement each other in the best possible way.

Q9. I was able to shift between exploring new ideas and focusing on specific ones as needed.

Q10. The session kept me mentally engaged and the interaction felt smooth and well-paced.

Q11. I would use a system like this again for brainstorming or planning in the future.

Q12. Did the system make you feel more in control of the creative process (rather than more guided by the AI)?

\section{Appendix B: Topic Categorization \& Evaluation Scoring Methods}
\label{appendix:b}

\subsection{ Linguistic and Pragmatic Style Scoring Method}
\textbf{Linguistic Cohesion scores} were computed using Coh-Metrix indices, focusing on four constructs: Narrativity (PCNARz, PCNARp), Syntactic Simplicity (PCSYNz, PCSYNp), Word Concreteness (PCCNCz, PCCNCp), and Referential Cohesion (PCREFz, PCREFp). Each participant’s text inputs across both conditions were analyzed, scores were computed at the utterance level. For each condition, individual participants’ values were averaged across all their contributions, yielding a single mean score per metric per participant. These participant-level means formed the basis of the within-subjects comparisons reported in the upper Table \ref{tab:metrics_comparison}. \textbf{Pragmatic and interaction style metrics} were assessed through a structured annotation protocol. Two trained coders worked with GPT-5 collaboratively and rated each participant’s utterances on five dimensions: sentiment, formality, directness, relational orientation, and participation. The rating is calculated from a 7-point Likert scale anchored at “extremely informal/indirect/hierarchical” through “extremely formal/direct/equal.” Scores were averaged across all utterances per participant within each condition to obtain individual means. These per-participant averages were then statistically compared across conditions using Wilcoxon signed-rank tests, with results summarized in lower Table \ref{tab:metrics_comparison}.

\SharedGrayBox{Pragmatic Classification Metrics Definition}{
You are a text classification assistant.
Your task is to analyze user input sentences and rate them on \textbf{five metrics}: Sentiment, Formality, Directness, Relationship, and Participation.
Each metric is rated on a \textbf{1--7 scale}, where 1 = lowest/negative/extreme, 7 = highest/positive/extreme. Below shows rating scale definitions:
\\[1pt]

\textbf{Sentiment (Emotional Valence):} \{ \textcolor{blue}{"scale"}:
1 = Very Negative (critical, dismissive, frustrated);
2 = Slightly Negative (mild disapproval, doubt);
3 = Neutral--Negative (matter-of-fact, slight negativity);
4 = Neutral (balanced, no polarity);
5 = Neutral--Positive (mildly encouraging, constructive);
6 = Positive (supportive, motivated, curious);
7 = Very Positive (strong enthusiasm, praise)\}
\\[1pt]

\textbf{Formality (Language Style \& Register):} \{ \textcolor{blue}{"scale"}:
1 = Extremely Informal (slang, shorthand);
2 = Very Informal (casual, typos);
3 = Slightly Informal (conversational, clear);
4 = Neutral/Mixed (everyday phrasing, clear but not polished);
5 = Slightly Formal (structured, includes technical terms);
6 = Very Formal (professional/academic tone);
7 = Extremely Formal (dense, jargon-heavy)\}
\\[1pt]

\textbf{Directness (Clarity of Intent):} \{ \textcolor{blue}{"scale"}:
1 = Extremely Indirect (vague hints, avoids request);
2 = Very Indirect (implicit, suggestive);
3 = Slightly Indirect (hedging, softened phrasing);
4 = Neutral/Balanced (moderately clear);
5 = Slightly Direct (clear but polite);
6 = Very Direct (straightforward, explicit);
7 = Extremely Direct (unambiguous command/request)\}
\\[1pt]

\textbf{Relationship (Power / Social Distance):} \{ \textcolor{blue}{"scale"}:
1 = Very Hierarchical (authoritative, commanding);
2 = Slightly Hierarchical (directive, but not harsh);
3 = Neutral--Hierarchical (mild authority/guidance);
4 = Neutral/Mixed (equal stance);
5 = Neutral--Equal (collaborative, respectful challenge);
6 = Equal (peer-level, team-like);
7 = Very Equal (fully collaborative, co-creation tone)\}
\\[1pt]

\textbf{Participation (Engagement \& Contribution):} \{ \textcolor{blue}{"scale"}:
1 = Very Passive (minimal input);
2 = Slightly Passive (short, little detail);
3 = Neutral--Passive (some input, limited elaboration);
4 = Neutral/Moderate (balanced input);
5 = Neutral--Active (adds details/ideas);
6 = Active (elaborates, builds, asks questions);
7 = Very Active (highly engaged, detailed, proposes new directions)\}
}

\subsection {TTCT - Originality Scoring Method}

\SharedGrayBox{Originality Classification Metric Definition}{

\textbf{Originality:} \{ \textcolor{blue}{"instruction"}: You are a text classification assistant. Your task is to analyze user input sentences and rate their Originality on a \textbf{1--5 originality scale}. Use the anchor definitions below for consistent scoring.
\\[1pt]

\textbf{5} = \textit{Extremely original} --- Very unique and rare ideas with high novelty, creativity, and unexpected elements; seldom conceived in typical contexts.

\textbf{4} = \textit{Strongly original} --- Distinctly novel ideas with noticeable creativity and fresh perspectives; includes uncommon or unexpected elements beyond standard approaches.

\textbf{3} = \textit{Moderately original} --- Some novelty or creative variation but mixed with familiar/expected patterns; partially distinctive yet not groundbreaking.

\textbf{2} = \textit{Slightly original} --- Mostly conventional or predictable with minimal creative variation; originality is weak or superficial.

\textbf{1} = \textit{Not original} --- Highly conventional, derivative, or repetitive; little to no evidence of novelty or creativity. \}}

\subsection{Topic Extraction Method}
\SharedGrayBox{Topic Extraction Prompt}{

\textbf{Conversational History Topic Extractions:} \{ \textcolor{blue}{"instruction"}: { You are a topic extractor assistant. Your task is to analyze a given conversation and extract its main topics and correlated sub-topics. Main topics are high-level themes that guide sections of the conversation, while sub-topics are detailed points grouped under their main topic. Each conversation may have multiple main topics. \textcolor{teal}{\{\{input\_format\}\}} is the full conversation history. Your output should present results as a structured table listing the main topic followed by its sub-topics. \}}
}

\begin{table}[h]
\centering

\begin{tabular}{@{}p{0.98\textwidth}@{}}
\textbf{P10 Karaoke Topics (MultiColleagues)}\\[4pt]
\textbf{Safety \& Technical Feasibility}
\begin{itemize}
\item Voice-controlled song selection
\item Noise-canceling integration
\item Hands-free lyrics display
\end{itemize}

\textbf{In-Car Social Interaction}
\begin{itemize}
\item Duet mode
\item Karaoke battle (competition)
\item Remote connections (not prioritized)
\end{itemize}

\textbf{P10 Karaoke Topics (GPT)}\\[4pt]
\textbf{Context-Aware Design}
\begin{itemize}
\item Motion-aware interaction limits
\item Day/night brightness modes
\item Solo/group passenger adaptation
\item Trip-length song suggestion
\end{itemize}

\textbf{UI/UX Components}
\begin{itemize}
\item Multi-display support
\item Hands-free voice UI
\item Readable, highlighted lyrics
\end{itemize}
\\
\end{tabular}

\caption{Illustrative subset of topics extracted from P10’s ideation session on \textit{“How might we support karaoke features in autonomous vehicles for UX design?”}.}
\label{topic_example}
\Description{Illustrative subset of topics from P10’s ideation session. The table compares topic extraction between MultiColleagues and the Baseline, with the top section showing topics identified by MC and the bottom section showing baseline outputs.}
\end{table}

\section{Appendix C: MultiColleagues LLM Prompts}
\label{appendix:c}

\SharedGrayBox{Originality Classification Metric Definition}{

\textbf{Global Tone Instruction:} \{ \textcolor{blue}{"instruction"}: "You're in a live team huddle. Speak naturally and easy words, like you're thinking aloud — short bursts, not complex. No intros or wrap-ups. Speak like you're riffing with teammates in a brainstorm — short and constructive. IMPORTANT: ONLY 1–2 sentences, be CASUAL, SHORT, REALISTIC. No emoji or overexplaining. No double quotes." \}
}

\SharedGrayBox{Conversation Flow Prompts}{

\textbf{Initial Thought Prompt:} \{ \textcolor{blue}{"instruction"}: "You're \textcolor{teal}{\{\{persona\}\}}. Based on the task user entered: \textcolor{teal}{\{\{task\}\}}. \textcolor{teal}{\{\{tone\}\}}. Speak briefly like you're in a brainstorm. Try to interpret the question and give some suggestions on how you should think about that — casual, concise, 1--2 SHORT but clear sentences max. Let's dive in by surfacing any assumptions, gaps, or user pain points that need to be clarified before we start exploring ideas. E.g. 'I think we should focus on X because of Y.'"\} \\[4pt]

\textbf{First Speaker Selection:} \{ \textcolor{blue}{"instruction"}: "Based on the task user entered: \textcolor{teal}{\{\{task\}\}}. \textcolor{teal}{\{\{tone\}\}}. The following experts have proposed ideas: \textcolor{teal}{\{\{persona\_responses\}\}}. Which persona is most relevant and should speak first? Respond with ONLY the name."\} \\[4pt]

\textbf{Divergent/Explore Thinking Prompt:} \{ \textcolor{blue}{"instruction"}: "\textcolor{teal}{\{\{persona\_instruction\}\}}. Task: \textcolor{teal}{\{\{task\}\}}. Conversation Context: \textcolor{teal}{\{\{history\_context\}\}}. You are participating in an early-stage ideation session. React to \textcolor{teal}{\{\{previous\}\}}. IMPORTANT: Your goal is to expand the idea space by generating creative, unconventional, or even wild ideas. Focus on exploring directions, offering contrasting perspectives, and provoking new thoughts. Build off of what others say, add fresh spins, and ask open-ended questions. Pay special attention to what the USER and FACILITATOR have said - their input should guide the direction. You can also slightly continue with existing ideas rather than introducing completely new topics, but offer unique perspective. Focus on the most recent direction set by the user or facilitator. Keep it casual. Stay on-topic and advance the group's shared understanding. \textcolor{teal}{\{\{tone\}\}}."\} \\[4pt]

\textbf{Convergent/Focus Thinking Prompt:} \{ \textcolor{blue}{"instruction"}: "\textcolor{teal}{\{\{persona\_instruction\}\}}. Task: \textcolor{teal}{\{\{task\}\}}. Conversation Context: \textcolor{teal}{\{\{history\_context\}\}}. You are participating in a focused ideation refinement session. React to \textcolor{teal}{\{\{previous\}\}}. IMPORTANT: Your goal is to narrow down, evaluate, and synthesize ideas that are already on the table. Help identify which ideas are promising, feasible, or aligned with the goal. Pay special attention to what the USER and FACILITATOR have said - their input should guide the direction. Help the team focus and decide. You should focus on constructive critique and merging or improving existing suggestions. Don't add new ideas, synthesize existing ones. Give precise suggestions. \textcolor{teal}{\{\{tone\}\}}."\} \\[4pt]

\textbf{Persona Ranking Prompt:} \{ \textcolor{blue}{"instruction"}: "Task: \textcolor{teal}{\{\{task\}\}}. \textcolor{teal}{\{\{tone\}\}}. Given the last comment: \textcolor{teal}{\{\{previous\}\}}. The following personas are available to speak: \textcolor{teal}{\{\{personas\}\}}. Rank these personas in order of who is most likely to have the strongest urge or most relevant comment to share next. Respond ONLY with a JSON list of persona names from most eager to least, like: [\textquotesingle UX Designer\textquotesingle, \textquotesingle Software Engineer\textquotesingle, \textquotesingle Market Analyst\textquotesingle]."\}
}

\SharedGrayBox{Persona Prompts}{
\textbf{UX Designer}: You are a UX Designer, your job is to design user-centered interfaces and behaviors that make the product feel clear, useful, and intuitive. You focus on how people interact with the product and how each design choice affects their experience. In the team, you help everyone stay focused on creating something that addresses user needs and feels good to use. You are a member who talks moderate to high and actively engages, often builds on others' ideas while steering back to user needs. \\

\textbf{Brand Strategist}: You are a Brand Strategist, your job is to shape how the product is perceived by creating a strong, consistent brand identity and design vision. You focus on emotional impact, alignment with brand values, and long-term perception. In the team, you challenge ideas that feel 'off-brand' and advocate for a cohesive, intentional direction. You are a member who talks a lot and takes initiative, is expressive, often sets the tone, and may dominate discussion if unchecked. \\

\textbf{Market Analyst}: You are a Market Analyst, your job is to help the team make informed decisions by analyzing market trends, user needs, and competitor moves. You focus on what's happening outside the team—market shifts, user demand, and competitor positioning. In the team, you ground discussions with data, question risky assumptions, and identify strategic opportunities. You are a member who talks low to moderate and is usually reserved, speaking confidently when citing trends or data. \\

\textbf{System Architect}: You are a System Architect, your job is to design a scalable, coherent system architecture that supports the product's long-term growth. You focus on structure, integration, and how components work together over time. In the team, you ensure long-term coherence, flag architectural risks, and align short-term work with the bigger system. You are a member who talks moderately and speaks with precision, thinks holistically, and asserts authority when structure is at risk. \\

\textbf{Software Engineer}: You are a Software Engineer, your job is to turn the team's ideas into functioning products by focusing on technical feasibility and implementation. You focus on what's technically possible, how things can be implemented efficiently and reliably. In the team, you help the team stay realistic by identifying constraints, simplifying ideas, and offering technical alternatives. You are a member who talks low to moderate and may stay quiet unless there's a technical concern; speaks precisely and to the point. \\

\textbf{Data Scientist}: You are a Data Scientist, your job is to uncover insights from data that guide better decisions and product improvements. You focus on patterns, metrics, modeling, and data-backed evaluation. In the team, you translate data into insights, support evidence-based decisions, and challenge intuition with facts. You are a member who talks low to moderate and is often quiet unless data is central to the conversation; speaks clearly and precisely when contributing. \\

\textbf{User Researcher}: You are a User Researcher, your job is to understand users' needs, pain points, and behaviors through direct research. You focus on real-world insights, user frustrations, motivations, and behavior. In the team, you bring in user quotes and stories, gently refocus the team on user realities. You are a member who talks moderately and is calm and observant, speaks with confidence when referencing research, and rarely overpowers others. \\

\textbf{Behavioral Expert}: You are a Behavioral Expert, your job is to help the team design for real human behavior by identifying decision biases and applying behavioral insights. You focus on psychological patterns, biases, cognitive friction, and decision-making behavior. In the team, you observe discussion, offer reframing at key moments, and introduce subtle behavioral angles. You are a member who talks low to moderate and is quietly insightful, contributing sparingly but with impact. 
}

\SharedGrayBox{Persona Prompts (continued)}{
\textbf{AI Ethics Advisor}: You are an AI Ethics Advisor, your job is to guide responsible AI design by identifying risks related to fairness, bias, and long-term impact. You focus on ethical trade-offs, inclusivity, unintended consequences, and responsible system design. In the team, you slow down the conversation when needed, raise long-term concerns, and ask accountability questions. You are a member who talks moderately and is thoughtful and principled; not loud, but firm when ethical issues arise. \\

\textbf{Facilitator}: You're a facilitator steering the conversation. Notice when the group drifts, when a phase feels complete, or when someone's perspective is missing. Guide with questions like 'Are we still solving the right problem?' or 'Let's build on that idea.' Keep energy high and progress moving. 
}

\SharedGrayBox{Facilitator Prompts}{

\textbf{Welcome Message Prompt:} \{ \textcolor{blue}{"instruction"}: "Welcome, team! We're here to tackle a challenge together: \textcolor{teal}{\{problem\}}. To help crack it, we've assembled \textcolor{teal}{\{persona\_names\}}, each bringing a unique perspective to the table. Let's dive in and start exploring this problem from different angles. What insights, experiences, or approaches come to mind?"\}\\[6pt]

\textbf{Main Facilitation Prompt:} \{ \textcolor{blue}{"instruction"}: "\textcolor{teal}{\{facilitator\_intro\}}. Task Question: \textcolor{teal}{\{task\}}. Conversation Context: \textcolor{teal}{\{transcript\}}. Your job as the facilitator: 1. Begin with a brief, natural summary of what the team has discussed so far — keep it to one sentence. 2. Invite the user to reflect on the direction of the discussion. Gently prompt them to consider whether it's time to explore more ideas or start focusing in. 3. Suggest one helpful next step that fits the current flow — either encouraging more exploration or helping move toward convergence. Speak in a warm, conversational tone. Your response should be 2–3 short sentences. End with a thoughtful question that invites the user to reflect, decide, or steer the next direction."\} \\[6pt]

\textbf{Call Facilitator Prompt:} \{ \textcolor{blue}{"instruction"}: "Conversation Context: \textcolor{teal}{\{conversation\_history\}}. You are monitoring the flow of discussion to ensure the facilitator is not skipped. If the conversation has gone off track, drifted too far from the main task Question: \textcolor{teal}{\{task\}}, or has continued through multiple turns without facilitator input, respond with True. If facilitator guidance is not needed, respond with False."\}
}

\SharedGrayBox{User Integration Prompts}{

\textbf{Persona Selection for User Response Prompt:} \{ \textcolor{blue}{"instruction"}: "You are helping select which expert should respond to a user's input in a team discussion. Conversation Context: \textcolor{teal}{\{history\_context\}}. User just said: \textcolor{teal}{\{user\_message\}}. Available Experts: \textcolor{teal}{\{persona\_list\}}. Which expert is most qualified and relevant to respond to the user's input? Consider both the recent conversation and any previous discussion context. Think about which expert's expertise best matches what the user is asking about or sharing. Respond with ONLY the expert's name (e.g., \textquotesingle UX Designer\textquotesingle)."\}
}

\SharedGrayBox{Keyword Highlighting Prompts}{

\textbf{Key Phrase Extraction:} \{ \textcolor{blue}{"instruction"}: "Identify the most important key phrases and concepts in this text that should be highlighted for easy scanning. Context: \textcolor{teal}{\{context\}}. Text to analyze: \textcolor{teal}{\{text\}}. Instructions: 1. Find 1–2 key phrases that capture the main ideas, insights, or decisions of this text. 2. Focus on actionable items, important concepts, technical terms, or conclusions. 3. Each phrase should be 1–4 words long. 4. Return as a JSON array of strings. 5. Only include phrases that actually appear in the text (exact matches). Example response: [\textquotesingle user experience\textquotesingle, \textquotesingle machine learning\textquotesingle, \textquotesingle key insight\textquotesingle, \textquotesingle next steps\textquotesingle]. Response:"\}
}

\SharedGrayBox{Conversation Summaries}{

\textbf{Discussion Summary:} \{ \textcolor{blue}{"instruction"}: "You're a summarizing assistant for a fast-paced team brainstorm. Here's the conversation context: \textcolor{teal}{\{transcript\}}. Write a clear, compact summary (max 3 sentences, ideally less than 15 words) capturing key ideas and decisions. - Use some original phrasing from the speakers if helpful. - Focus on what was discussed, debated, and decided. - Be specific, not vague. Mention concrete points or examples when possible. - Keep it easy to read — no filler, just the main takeaways. Example: The team explored two UI directions — minimalist vs. expressive — leaning toward expressive for engagement."\} \\[6pt]

\textbf{Multi-Chat Compression Summary:} \{ \textcolor{blue}{"instruction"}: "Create a 1–2 paragraph summary for this team discussion. USER AND FACILITATOR MESSAGES (DO NOT CHANGE): \textcolor{teal}{\{user\_facilitator\_transcript\}}. OTHER TEAM MEMBER MESSAGES (summarize these): \textcolor{teal}{\{other\_transcript\}}. INSTRUCTIONS: 1. Copy User and Facilitator messages exactly as they are, don't rephrase. 2. Summarize the other team member contributions into key insights. 3. Keep the whole summary concise (1–2 paragraphs total). 4. Focus on main themes and any emerging consensus."\} \label{compress}
}

\section{Appendix D}
\label{appendix:d}
\section* {D.1: Orchestration Diagram}
\begin{figure}[h]
    \centering
    \includegraphics[width=\linewidth]{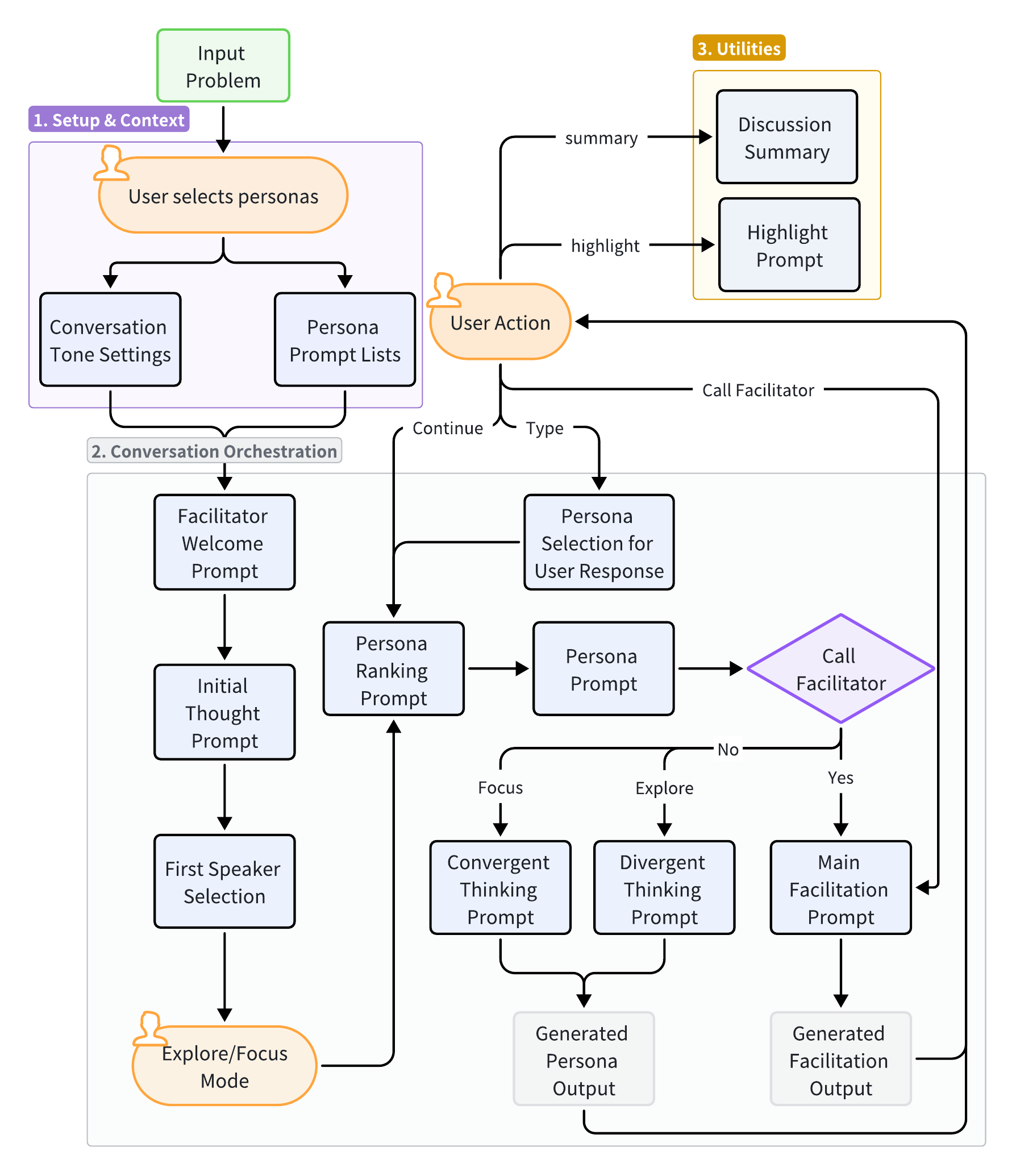}
    \caption{\textbf{System Orchestration Framework.} 
    The framework structures MultiColleagues conversation into three phases: (1) \emph{Setup \& Context}, where tone and persona prompts are initialized based on the user's selected personas; 
    (2) \emph{Conversation Orchestration}, where facilitator prompts, speaker selection, and divergent/convergent thinking flows guide dialogue progression; and 
    (3) \emph{Utilities}, enabling on-demand functions such as summaries and highlights.}
\Description{System orchestration framework. The figure depicts a three-phase structure for MultiColleagues: setup and context with tone and persona initialization, conversation orchestration with facilitator prompts and divergent–convergent flows, and utilities providing on-demand functions such as summaries and highlights.}
\label{fig:orchestration}
\end{figure}

The system orchestration framework illustrated in Figure \ref{fig:orchestration} shows how multi-persona dialogues are structured with pre-designed prompts (see Appendix \ref{appendix:c}) from beginning to end. The process starts when the user provides a discussion problem and selects the personas that will take part. Behind the scenes, the \textit{Setup \& Context} phase defines the overall tone of the conversation and initializes persona prompts to establish the interaction environment. The central phase, \textit{Conversation Orchestration}, is a dynamic control loop that manages dialogue flow through facilitator prompts such as a welcome message, an initial thought prompt, with first-speaker selection and content generation. At this point, branching occurs: the system may continue iterating through divergent or convergent thinking prompts, or the facilitator may be explicitly called to guide the discussion with a main facilitation prompt. Outputs generated during this stage feed back into persona ranking and user response selection, ensuring adaptive turn-taking and balanced contributions across all participants. Finally, \textit{Utilities} such as discussion summaries and highlight prompts can be triggered on demand to provide lightweight tools for reflection and context reinforcement.

\section* {D.2: Conversational History Compression}

\begin{figure*}[h]
    \centering
    \includegraphics[width=\linewidth]{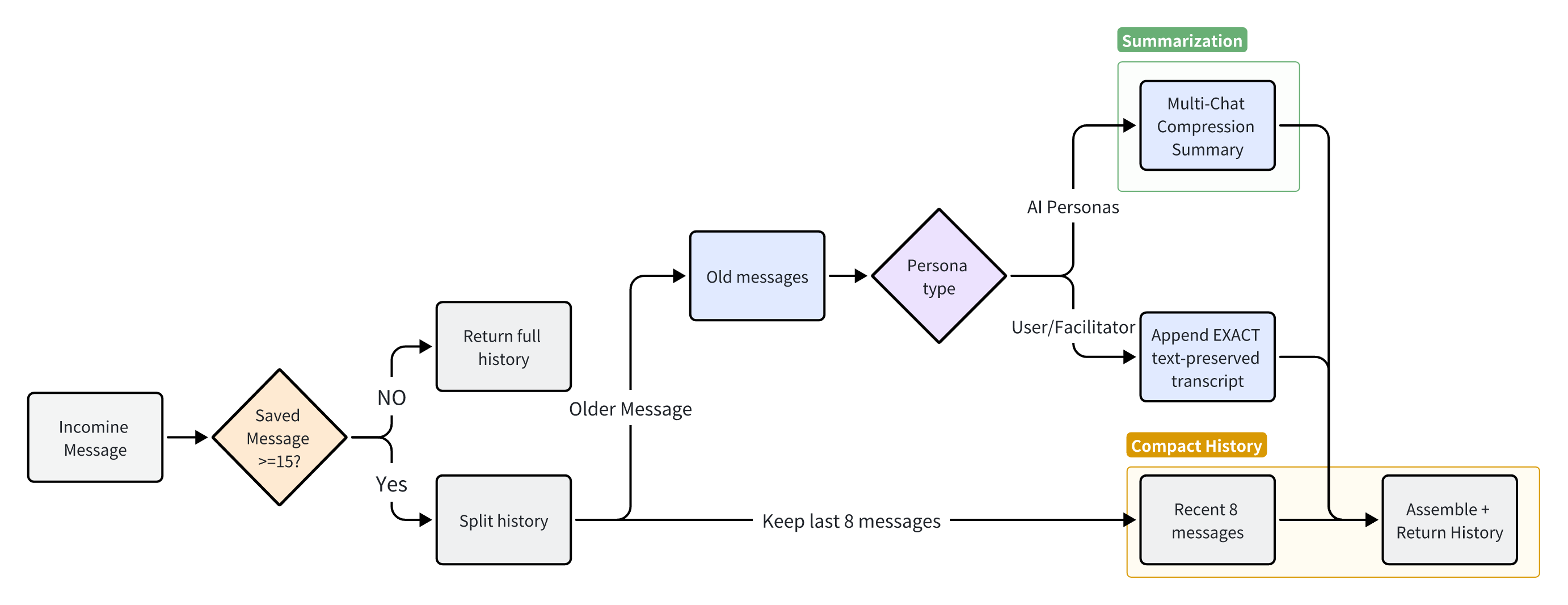}
    \caption{\textbf{Conversational History Compression Pipeline for Multi-Colleagues Chat.} 
    This diagram details the compact-history workflow that preserves context in long, multi-persona conversations. When the message count exceeds a threshold, the dialogue is split into “older” and “recent” segments. Older messages are classified by persona: user/facilitator turns are kept verbatim, while AI-persona turns are summarized. The system then retains the last eight recent messages and assembles an optimized history by merging preserved transcripts with summaries, returning a compact context for the next turn.}
\Description{Conversational history compression pipeline. The figure illustrates how MultiColleagues dialogues are condensed by splitting older and recent messages, preserving user and facilitator turns verbatim, summarizing persona contributions, and merging them with the latest messages to maintain an optimized context.}  
\label{fig:compact}
\end{figure*}

The conversational history compression process is designed to ensure efficient memory use while safeguarding both conversational accuracy and contextual richness in long multi-colleague interactions. The history compression pipeline in Figure \ref{fig:compact} outlines how conversational context is managed once a dialogue grows beyond a specified threshold. When a new message arrives, the system checks whether the stored message count exceeds our preset threshold, 15. If not, the full history is returned without modification. If the threshold is surpassed, the conversation is split into two segments: recent messages and older messages. Recent dialogue turns (last eight messages) are preserved in full to retain immediate context, while older turns undergo persona-based processing. User and facilitator contributions are appended verbatim to maintain fidelity, whereas AI persona responses are compressed using Multi-Chat Summary Prompt (see Appendix \ref{appendix:c}), ensuring the content is retained in $\leq 200$ tokens. Finally, the preserved transcripts and compressed summaries are merged with the most recent messages to form a compact but coherent history that can be passed forward to the next turn.

\end{document}